\documentclass[11pt,a4paper]{article}
\usepackage{amsmath}
\usepackage[mathcal]{eucal}
\usepackage{latexsym}
\usepackage{amssymb}
\begin{titlepage}
\title{Instanton representation of Plebanski gravity: XIII. Canonical structure of the Petrov classification of nondegenerate spacetimes}
\author{Eyo Eyo Ita III}
\medskip
\input amssym.def
\input amssym.tex

\def \in{\indent}

\begin{document}
\maketitle
\bigskip
\centerline{Department of Applied Mathematics and Theoretical Physics} 
\smallskip
\centerline{Centre for Mathematical Sciences, University of Cambridge, Wilberforce Road}
\smallskip
\centerline{Cambridge CB3 0WA, United Kingdom}
\smallskip
\centerline{eei20@cam.ac.uk} 
 
\bigskip

\begin{abstract}
The instanton representation of Plebanski gravity admits a natural canonical structure where the (densitized) eigenvalues of the CDJ matrix are the basic momentum space variables.  Canonically conjugate configuration variables exist for six distinct configurations in the full theory, referred to as quantizable configurations.  The CDJ matrix relates to the Petrov classification and principal null directions of spacetime, which we directly correlate to these quantizable degrees of freedom.  The implication of this result is the ability to perform a quantization procedure for spacetimes of Petrov Type I, D, and O, using the instanton representation.
\end{abstract}
\end{titlepage}

\section{Introduction}

In the full theory of general relativity there currently remain at least three unresolved questions.  (i) One question is to find the projection from the full unreduced phase space $\Omega$ of the theory to the 
physical phase space $\Omega_{Phys}$, through implementation of the initial value constraints.  (ii) The second main question regards the quantization of the theory, which has posed technical difficulties in various approaches.  (iii) The third main question is the verification of the quantum theory in terms of quantities which can be measured in the classical limit.  The aim of this paper is to provide a preliminary addressal of these questions using the instanton represetation of Plebanski gravity (See e.g. paper II and listings therein).  The CDJ matrix $\Psi_{ae}$ is a $SO(3,C)\otimes{SO}(3,C)$-valued matrix which was introduced \cite{CAP} in order to construct a solution to the diffeomorphism and the Hamiltonian constraints in Ashtekar variables.  The instanton representation uses $\Psi_{ae}$ as the fundamental momentum space variable.  It so happens that the initial value constraints of GR when 
written on $\Omega_{Inst}=(\Psi_{ae},A^a_i)$, the phase space of the instanton representation, are essentially constraints on $\Psi_{ae}$.\footnote{Here $A^a_i$ is the self-dual Ashtekar connection, which is the configuration space variable.  By convention lowercase symbols from the beginning of the Latin alphabet $a,b,c,\dots$ signify internal $SO(3,C)$ indices, while those from the middle $i,j,k,\dots$ signify spatial indices.}  This feature enables one to readily address question (i) by projecting directly to the reduced momentum space, where it remains to find the physical principle fixing the canonically conjugate configuration variables in the preservation of a cotangent bundle structure on this reduced space.\par
\indent
Further investigation of the physical interpretation of the CDJ matrix $\Psi_{ae}$ reveals that it correlates to the algebraic properties of spacetime which are independent of coordinates and of tetrad frames.  In the addressal of question (iii) above, it then suffices to correlate these aspects of spacetime to degrees of freedom which in the instanton representation can be quantized.  We show in this paper that such degrees of freedom correspond to the nondegenerate spacetimes, those spacetimes whose self-dual Weyl curvature tensor possess three linearly independent eigenvectors.  The eigenvalues of the CDJ matrix for these spacetimes encode their Petrov classification and certain information pertaining to the principal null directions (PND).  It is then clear, in the addressal of question (ii) above, that one may formulate a quantum theory of the CDJ matrix which admits a direct link to these PND which are in principle directly measurable in the classical limit.\par
\indent
A closer analysis of the canonical structure of the instanton representation indicates a potential obstruction in that there are no configuration space variables canonically conjugate to the bare CDJ matrix $\Psi_{ae}$.\footnote{This is with the exception of the trace of $\Psi_{ae}$, whose canonically conjugate variable is $L_{CS}$, the spatial Chern--Simons Lagrangian.}  This is a consequence of the fact that the transformaton from the Ashtekar variables into the instanton representation is a noncanonical transformation.  On first sight, it may seem that this prevents one from formulating a quantum theory where $\Psi_{ae}$ is the momentum space variable.  However, we have found that this obstruction is circumvented by using a densitized CDJ matrix $\widetilde{\Psi}_{ae}=\Psi_{ae}(\hbox{det}A)$ in lieu of $\Psi_{ae}$ as the basic momentum space variable, and projecting to the kinematic 
phase space $\Omega_{Kin}$.\footnote{This is defined as the phase space of the instanton representation after implementation of the Gauss' law and diffeomorphism constraints, and prior to the Hamiltonian constraint.}  This results in six distinct quantizable configurations of the instanton representation, which correlate to Petrov classifications in the classical limit.\par  
\indent  
The organization of this paper is as follows.  Section 2 transforms vacuum GR from the Ashtekar variables directly into the instanton representation using the CDJ Ansatz, which holds on the space of 
nondegenerate variables.\footnote{As shown in Paper II, the instanton representation can also be obtained by implementing the simplicity constraint and eliminating the densitized triad $\widetilde{\sigma}^i_a$ directly from the 
starting Plebanski action.}  Additionally, we show how the CDJ matrix facilitates the implementation of the initial value constraints, and how the eigenvalues of $\Psi_{ae}$ emerge as a natural candidate for the momentum space variables.  In section 3 we delineate the conditions for which there exist globally holonomic coordinates on the instanton representation kinematic configuration space $\Gamma_{Inst}$ which are canonically conjugate to these (densitized) eigenvalues.  This limits one to nondegenerate connections $A^a_i$ with three degrees of freedom per point, of which there are six distinct configurations corresponding to nondegenerate metrics.\footnote{Inherently, this limits the results of this paper to quantization to spacetimes of Petrov type $I$, $D$ and $O$, where the CDJ matrix is diagonalizable.}  Section 4 elucidates the natural correspondence from these quantizable configurations to the instrinsic $SO(3,C)$ frame, the frame corresponding to the kinematic phase space.  While the instanton representation is not canonically related to the Ashtekar variables on the full starting phase space $\Omega_{Inst}$, we show that it is canonically related in the intrinsic $SO(3,C)$ frame.  This establishes the reduced phase space of the Ashtekar variables, which corresponds to nondegenerate metrics.  Section 5 creates a library of the so-called `quantizable' instanton representation configurations, by explicit construction.\par
\indent
Having demonstrated that the eigenvalues of $\Psi_{ae}$ admit a quantization of the full theory consistent with the implementation of the initial value constraints, we now show the manner in which these eigenvalues directly correlate to aspects of spacetime which are directly measurable.  In this section there is some background which is provided regarding the Weyl curvature tensor and its two-component spinor formalism and principal null directions.  We then explicitly relate these quantities to the CDJ matrix, which establishes the direct link from the classical to the quantum theory.  In this paper we establish the canonical structure required for quantization.  The full quantization procedures, including the Hilbert space structure, is reserved for separate papers.

\newpage

\section{From the Ashtekar variables into the instanton representation}

\noindent
In the Ashtekar description of gravity the basic phase space variables are a self-dual $SU(2)_{-}$ connection and a densitized triad $(A^a_i,\widetilde{\sigma}^i_a)\in\Omega$.\footnote{By our convention, lowercase symbols from the beginning part of the Latin alphabet $a,b,c,\dots$ signify internal $SO(3,C)$ indices, while from the middle of the alphabet $i,j,k,\dots$ signify spatial indices in 3-space $\Sigma$.}  The Ashtekar connection is given by

\begin{eqnarray}
\label{ASHTEKARCONNECTION}
A^a_i=\Gamma^a_i+\beta{K}^a_i,
\end{eqnarray}

\noindent
where $\Gamma^a_i$ is the spin connection compatible with the triad defined 
by $\widetilde{\sigma}^i_a$, $\beta$ is the Immirzi parameter, which we choose to be $-i$, and $K^a_i$ is the triadic form of the extrinsic curvature of 3-space $\Sigma$.  The 3+1 decomposition of the resulting action for vacuum general relativity in $M=\Sigma\times{R}$, where $M$ is a 4-dimensional spacetime manifold foliated by 3-dimensional spatial hypersurfaces $\Sigma$, is given by

\begin{eqnarray}
\label{ASHACTION}
I_{Ash}=\int{dt}\int_{\Sigma}\widetilde{\sigma}^i_a\dot{A}^a_i-A^a_0G_a-N^iH_i-\underline{N}H.
\end{eqnarray}

\noindent
Equation (\ref{ASHACTION}) is a canonical one form $\boldsymbol{\theta}_{Ash}$ minus a linear combination of first class constraints smeared by 
auxilliary fields \cite{ASH1},\cite{ASH2},\cite{ASH3}.  The auxilliary fields are $N^i$, $A^a_0$ and $\underline{N}=N(\hbox{det}\widetilde{\sigma})^{-1/2}$, respectively the shift vector, $SO(3,C)$ rotation angle and lapse density function,\footnote{For $N$ real the action (\ref{ASHACTION}) corresponds to a spacetime of Euclidean signature.  For Lorentzian signature one may perform a Wick rotation $N\rightarrow{i}N$.} and the 
corresponding constraints $H_i$, $G_a$ and $H$ are the diffeomorphism, Gauss' law and Hamiltonian constraints.  The diffeomorphism constraint is given by

\begin{eqnarray}
\label{DEEEF}
H_i=\epsilon_{ijk}\widetilde{\sigma}^j_aB^k_a=0,
\end{eqnarray}

\noindent
which signifies invariance under spatial diffeomorphisms in $\Sigma$.  The Hamiltonian constraint signifies invariance under deformations normal to $\Sigma$ and is given by

\begin{eqnarray}
\label{HAAAMMU}
H={\Lambda \over 3}\epsilon_{ijk}\epsilon^{abc}\widetilde{\sigma}^i_a\widetilde{\sigma}^j_b\widetilde{\sigma}^k_c+\epsilon_{ijk}\epsilon^{abc}\widetilde{\sigma}^i_a\widetilde{\sigma}^j_bB^k_c=0,
\end{eqnarray}

\noindent
where $\Lambda$ is the cosmological constant and $B^i_a={1 \over 2}\epsilon_{ijk}F^a_{jk}$ is the magnetic field derived from the curvature of the Ashtekar connection $A^a_i$, where

\begin{eqnarray}
\label{CURVATOORE}
F^a_{ij}=\partial_iA^a_j-\partial_jA^a_i+f^{abc}A^b_iA^c_j.
\end{eqnarray}

\noindent
The Gauss' law constraint, which signifies invariance under left-handed $SO(3,C)$ rotations on internal indices, is given by 

\begin{eqnarray}
\label{GGAAUUSS}
G_a=D_i\widetilde{\sigma}^i_a=\partial_i\widetilde{\sigma}^i_a+f_{abc}A^b_i\widetilde{\sigma}^i_c=0,
\end{eqnarray}

\noindent
where $f_{abc}$ are the structure constants for $SO(3,C)$.\par  
\indent
In this paper we would like to find a set of degrees of freedom suitable for quantizaiton, which entails an implementation of the initial value 
constraints (\ref{DEEEF}), (\ref{HAAAMMU}) and (\ref{GGAAUUSS}).  Let us transform (\ref{ASHACTION}) into a new set of variables variables using the CDJ Ansatz

\begin{eqnarray}
\label{CEEDEEJAY}
\widetilde{\sigma}^i_a=\Psi_{ae}B^i_e
\end{eqnarray}

\noindent
attributed to Riccardo Capovilla, John Dell and Ted Jacobson \cite{CAP}, where $\Psi_{ae}\in{SO}(3,C)\otimes{SO}(3,C)$ is the CDJ matrix.\footnote{We actually use the inverse of the matrix used in \cite{CAP}, and allow for a nonzero trace.}  Equation (\ref{CEEDEEJAY}) is good as long as $B^i_a$ is nondegenerate, and has been shown in \cite{CAP} to allow explicit solution of $H_{\mu}=(H,H_i)$ algebraically at the classical level.  Under (\ref{CEEDEEJAY}) the 
action (\ref{ASHACTION}) becomes

\begin{eqnarray}
\label{CEEDEEACTION}
I_{Inst}=\int{dt}\int_{\Sigma}d^3x\Psi_{ae}B^i_e\dot{A}^a_i-\bigl(\epsilon_{ijk}N^iB^j_aB^k_e+{A}^a_0\textbf{w}_e\bigr)\Psi_{ae}\nonumber\\
+N(\hbox{det}B)^{1/2}\sqrt{\hbox{det}\Psi}\bigl(\Lambda+\hbox{tr}\Psi^{-1}\bigr).
\end{eqnarray}

\indent
We have also defined

\begin{eqnarray}
\label{VECFIELD}
\textbf{w}_e\{\Psi_{ae}\}=B^i_e\partial_i\Psi_{ae}+\bigl(f_{ghe}\delta_{af}+f_{fae}\delta_{gh}\bigr)C_{eh}\Psi_{fg}
\equiv\textbf{w}_a^{fg}\{\Psi_{fg}\},
\end{eqnarray}

\noindent
which comes from the transformation of the Gauss' law constraint

\begin{eqnarray}
\label{VECFIELD1}
G_a=D_i\widetilde{\sigma}^i_a=D_i(\Psi_{ae}B^i_e)=\Psi_{ae}D_iB^i_e+B^i_eD_i\Psi_{ae},
\end{eqnarray}

\noindent
where $C_{ae}=A^a_iB^i_e$ is the magnetic helicity density (See e.g. Paper VI).  Upon use of the Bianchi identity $D_iB^i_e=0$, the definition (\ref{VECFIELD}) follows from the evaluation of the covariant derivative $D_i\Psi_{ae}$ in (\ref{VECFIELD1}) in the tensor representation of the gauge group.\par
\indent

\subsection{Consistency with the algebraic constraints}

\noindent
Let us now demonstrate consistency of the CDJ Ansatz (\ref{CEEDEEJAY}) with the initial value constraints, from a different approach to that introduced in \cite{CAP}.  The CDJ matrix $\Psi_{ae}$ can be parametrized by its symmetric and antisymmetric parts, $\lambda_{ae}$ and $a_{ae}$ respectively, which can in turn be parametrized by a polar decomposition

\begin{eqnarray}
\label{SOLU}
\Psi_{ae}=a_{ae}+\lambda_{ae}=\epsilon_{aed}\psi_d+O_{af}(\vec{\theta})\lambda_fO^T_{fe}(\vec{\theta}),
\end{eqnarray}

\noindent
where $\psi_d=\epsilon_{dae}\Psi_{ae}$ is a $SO(3,C)$ 3-vector.\footnote{This decomposition is possible when $\Psi_{ae}$ contains three linearly independent eigenvectors, which as we will see restricts one to spacetimes of Petrov type $I$, $D$ and $O$.}  Here, $\vec{\lambda}\equiv(\lambda_1,\lambda_2,\lambda_3)$ are the eigenvalues of $\lambda_{ae}=\Psi_{(ae)}$, while $O_{ae}\in{SO}(3,C)$ implements a complex orthogonal transformation 
of $\vec{\lambda}$ parametrized by three complex angles $\vec{\theta}=(\theta_1,\theta_2,\theta_3)$.  In exponential form 
this is given by $O=e^{\theta\cdot{T}}$, where $T$ are generators satisfying the $SO(3)$ Lie algebra

\begin{eqnarray}
\label{ALGEEB}
[T_f,T_g]=i\epsilon_{fgh}T_h.
\end{eqnarray}

\noindent
The diffeomorphism constraint in the instanton representation is given by

\begin{eqnarray}
\label{DIFF}
H_i=\epsilon_{ijk}B^j_aB^k_e\Psi_{ae}=\epsilon_{ijk}B^j_aB^k_e\epsilon_{aed}\psi_d=0~~\forall~x,
\end{eqnarray}

\noindent
which implies that the antisymmetric part of the CDJ matrix vanishes ($\psi_d=0$), or that CDJ matrix is symmetric $\Psi_{ab}=\Psi_{(ab)}$.  The Hamiltonian constraint is given by the last term 
of (\ref{CEEDEEACTION}).  Since $(\hbox{det}B\neq{0}$ and $(\hbox{det}\Psi\neq{0})$, then it suffices that

\begin{eqnarray}
\label{HAAAAAMM}
{1 \over 2}Var\Psi+\Lambda\hbox{det}\Psi=0~~\forall{x}.
\end{eqnarray}

\noindent
Substitution of the parametrization (\ref{SOLU}) into (\ref{HAAAAAMM}) after dividing by $\hbox{det}B\neq{0}$, which requires nondegeneracy of $B^i_a$, yields

\begin{eqnarray}
\label{HOME}
H={1 \over 2}Var(\Psi_{(ae)})+\Lambda\hbox{det}(\Psi_{(ae)})+(\Lambda\Psi_{(ae)}-\delta_{ae})\psi_a\psi_e=0.
\end{eqnarray}

\noindent
The first two terms of (\ref{HOME}) can be rewritten explicitly in terms of the eigenvalues of $\lambda_{ae}$ due to the cyclic property of the trace, yielding\footnote{See Appendix A for the details of the derivation.}

\begin{eqnarray}
\label{HOME1}
H=\bigl(\lambda_1\lambda_2+\lambda_2\lambda_3+\lambda_3\lambda_1\bigr)+\Lambda\lambda_1\lambda_2\lambda_3\nonumber\\
+\bigl(\Lambda{O}_{af}O_{ef}\lambda_f-\delta_{ae}\bigr)\psi_a\psi_e=0.
\end{eqnarray}

\noindent
Using $\psi_d=0$ from (\ref{DIFF}), on the space of solutions to $H_i$ for nondegenerate $B^i_a$, the terms in (\ref{HOME1}) quadratic in $\psi_d$ vanish along with the $SO(3,C)$ 
matrix $O_{ae}$.  The Hamiltonian constraint (\ref{HOME1}) then reduces to the following algebraic relation amongst the eigenvalues $\lambda_f$

\begin{eqnarray}
\label{SOLU1}
\Lambda+{1 \over {\lambda_1}}+{1 \over {\lambda_2}}+{1 \over {\lambda_3}}=0\rightarrow\lambda_3=-\Bigl({{\lambda_1\lambda_2} \over {\Lambda\lambda_1\lambda_2+\lambda_1+\lambda_2}}\Bigr).
\end{eqnarray}

\noindent 
Hence for $\lambda_f\neq{0}$ (\ref{SOLU1}) fixes one eigenvalue $\lambda_3$ completely in terms of the remaining two eigenvalues $\lambda_1$ and $\lambda_2$, with no appearance of the $SO(3,C)$ angles $\vec{\theta}$.  Observe that the implementation of $H_{\mu}=(H,H_i)$ has resulted in a reduction of $\Psi_{ae}$ by four D.O.F. to its eigenvalues with no corresponding restriction on 
the Ashtekar connection $A^a_i$.\par
\indent  
We will ultimately choose $\lambda_1$ and $\lambda_2$ as the physical D.O.F. for the momentum space of the instanton representation.  To endow the theory with symplectic structure, we must find two D.O.F. corresponding to the configuration space variables $\Gamma_{Phys}$ canonically conjugate to $(\lambda_1,\lambda_2)$, which will be one of the main results of this paper.

\indent

\subsection{Consistency with the Gauss' law constraint}

While the $SO(3,C)$ matrix $O_{ae}[\vec{\theta}]$ has been eliminated from the Hamiltonian constraint $H$, owing to its restriction to the invariants of $\Psi_{(ae)}$ on the diffeomorphism constraint shell, it will appear explicitly in the Gauss' law constraint $G_a$.  The unconstrained momentum space D.O.F. have already been reduced to $\vec{\lambda}=(\lambda_1,\lambda_2,\lambda_3)$ at the level prior to implementation of $H$.  Therefore $G_a$ should not reduce these particular D.O.F. any further, and neither does it impose any restrictions on $A^a_i\in\Gamma_{Inst}$.  Hence, $G_a$ must be viewed as a constraint on the angles $\vec{\theta}$, which in turn define a special $SO(3,C)$ frame.   The eigenvalues $\lambda_f$ must then be rotated from the intrinsic frame where $\vec{\theta}=0$ into the $SO(3,C)$ frame where $\Psi_{ae}$ becomes annihilated by $G_a$.  This frame, fixed by the correctly chosen $\vec{\theta}$, should correspond to a solution to the initial value constraints of GR.\par
\indent
The Gauss' law constraint is given by

\begin{eqnarray}
\label{GAU}
\textbf{w}_e\{\Psi_{ae}\}=B^i_eD_i\Psi_{ae}=\textbf{v}_e\{\Psi_{ae}\}+\bigl(f_{abf}\delta_{ge}+f_{ebg}\delta_{af}\bigr)C_{be}\Psi_{fg}=0
\end{eqnarray}

\noindent
where $\textbf{v}_e=B^i_e\partial_i$ and $C_{be}=A^b_iB^i_e$ is defined as the helicity density matrix.  Unlike the diffeomorphism and Hamiltonian constraints which are algebraic, the Gauss' law constraint is a set of differential equations.  To solve (\ref{GAU}) one may first decompose $\Psi_{ae}$ into a basis of shear (off-diagonal symmetric) and anisotropy (diagonal) elements, using the Cartesian representation

\begin{eqnarray}
\label{CARTE}
\Psi_{ae}=(e^f)_{ae}\varphi_f+(E^f)_{ae}\Psi_f,
\end{eqnarray}

\noindent
where we have defined

\begin{displaymath}
E^1_{ae}=
\left(\begin{array}{ccc}
0 & 0 & 0\\
0 & 0 & 1\\
0 & 1 & 0\\
\end{array}\right);~~
E^2_{ae}=
\left(\begin{array}{ccc}
0 & 0 & 1\\
0 & 0 & 0\\
1 & 0 & 0\\
\end{array}\right);~~
E^3_{ae}=
\left(\begin{array}{ccc}
0 & 1 & 0\\
1 & 0 & 0\\
0 & 0 & 0\\
\end{array}\right),
\end{displaymath}

\begin{displaymath}
e^1_{ae}=
\left(\begin{array}{ccc}
1 & 0 & 0\\
0 & 0 & 0\\
0 & 0 & 0\\
\end{array}\right);~
e^2_{ae}=
\left(\begin{array}{ccc}
0 & 0 & 0\\
0 & 1 & 0\\
0 & 0 & 0\\
\end{array}\right);~
e^3_{ae}=
\left(\begin{array}{ccc}
0 & 0 & 0\\
0 & 0 & 0\\
0 & 0 & 1\\
\end{array}\right)
.
\end{displaymath}

\noindent
The CDJ matrix (\ref{CARTE}) in matrix form is given by

\begin{displaymath}
\Psi_{ae}=
\left(\begin{array}{ccc}
\varphi_1 & \Psi_3 & \Psi_2\\
\Psi_3 & \varphi_2 & \Psi_1\\
\Psi_2 & \Psi_1 & \varphi_3\\
\end{array}\right)
.
\end{displaymath}

\noindent
The Gauss' law constraint then becomes

\begin{eqnarray}
\label{CARTE1}
(e^f)_{ae}\textbf{w}_e\{\varphi_f\}+(E^f)_{ae}\textbf{w}_e\{\Psi_f\}
\end{eqnarray}

\noindent
which equivalently is given by $\Psi_f=\hat{J}_f^g\varphi_g$.  We have defined the Gauss' law propagator $\hat{J}_f^g$ from the anisotropy to the shear elements, given by

\begin{eqnarray}
\label{CARTE2}
\hat{J}_f^g=-\bigl((E^f)_{ae}\textbf{w}_e\bigr)^{-1}(e^g)_{ab}\textbf{w}_b.
\end{eqnarray}

\noindent
For the purposes of the present paper it suffices to note that $G_a$ reduces $\Psi_{ae}$ by three unphysical D.O.F. $\Psi_f$, leaving remaining $\varphi_f$.\footnote{The details of the inversion (\ref{CARTE2}), as well as the explicit solution algorithm for the angles $\vec{\theta}$, are treated in Papers VI, VII and VIII.  The idea is that one must solve (\ref{GAU}) explicitly for $\Psi_f=\vec{\Psi}[\vec{\varphi}]$ for each configuration $A^a_i\in\Gamma$.  Certain configurations with yield well-defined $\vec{\theta}$ and other configurations will not.  But whatever the configuration chosen, there exists a map from the physical degrees of 
freedom $\vec{\lambda}$ to the angles $\vec{\theta}$.}  Within this context, $G_a$ establishes a map from the physical D.O.F. $\vec{\lambda}$ to the angles

\begin{eqnarray}
\label{MAPPP}
\lambda_f\longrightarrow\varphi_f\longrightarrow\bigl(\Psi_f[\vec{\lambda};A^a_i],\vec{\theta}[\vec{\lambda};A^a_i]\bigr).
\end{eqnarray}

\noindent
First one chooses a particular  configuration for the Ashtekar connection $A^a_i$ and then finds the Gauss' law propagator (\ref{CARTE2}) corresponding to that configuration $\hat{J}_f^g=\hat{J}_f^g[A]$.  To find the 
angles $\vec{\theta}$, one then equates the polar representation of $\Psi_{ae}$ to the Cartesian representation

\begin{displaymath}
\lambda_{ab}=O_{ae}(\vec{\theta})
\left(\begin{array}{ccc}
\lambda_1 & 0 & 0\\
0 & \lambda_2 & 0\\
0 & 0 & \lambda_3\\
\end{array}\right)
_{ef}O^T_{fb}(\vec{\theta})
=\bigl((e^g)_{ae}+(E^f)_{ae}\hat{J}_f^g\bigr)\varphi_g
\end{displaymath}

\noindent
and then solves for $\vec{\theta}$ explicitly in terms of the eigenvalues $\lambda_f$ and $A^a_i$.  One then repeats the procedure for all configurations $A^a_i$, which defines a 
functional $\vec{\theta}=\vec{\theta}[\lambda_f;A^a_i]$ constituting a solution space for $G_a$.  Another method is to write the Gauss' law constraint as a set of differential equations directly on the angles $\vec{\theta}$, thus bypassing the Cartesian representation, as shown in Paper VIII.  Note that finding $\Psi_{ae}\in{Ker}\{G_a\}$ does not place any restriction on $A^a_i\in\Gamma$.\footnote{This is becuase $A^a_i$ is one of the inputs into Gauss' law constraint.  Certain configurations will yield a well-defined solution and other configurations will not.}  To obtain a symplectic structure on the reduced phase space under $G_a$ we must find the appropriate 
restriction required of $A^a_i$ by alternate means, which brings us to the issue of the existence of holonomic coordinates.

\newpage

\section{Globally holonomic configuration space for the instanton representation}

\noindent
When the initial value constraints have been implemented, the phase space of GR should consist of four degrees of freedom per point.\footnote{This refers to complex phase space degrees of freedom.  The implementation of reality conditions is a separate procedure from the initial value constraints.}  Denote the kinematic phase space $\Omega_{Kin}$ as the phase space at the level where the diffeomorphism and the Gauss' law constraints, but not the Hamiltonian constraint, have been implemented.  At this stage $\Omega_{Kin}$ should consist of six D.O.F. per point and we would like to use the three eigenvalues $\lambda_f$ as the momentum space part of these degrees of freedom.  Were this to be the case then the canonically conjugate configuration space variables should be determined, if they exist, so that a quantization procedure can be carried out.\par
\indent
The canonical structure of (\ref{CEEDEEACTION}) suggests naively that $\Psi_{ae}$ should be canonically conjugate to a variable $X^{ae}$ whose velocity is $\dot{X}^{ae}=B^i_e\dot{A}^a_i$.\footnote{The variations 
$\delta{X}^{ae}=B^i_e\delta{A}^a_i\in{T}^{*}(\Gamma)$ are well-defined in the cotangent space to configuration space $\Gamma$ (See e.g. \cite{SOO} and \cite{SOO1}).  The variables $X^{ae}$ were first 
discovered by Chopin Soo in \cite{SOO}, due to their natural adaptability to the gauge invariances of GR.}  However, $X^{ae}$ does not exist globally as a holonomic coordinate on $\Gamma_{Inst}$ for arbitrary $A^a_i$, which by our interpretation constitutes an obstruction to quantization.  Equation (\ref{CEEDEEJAY}) transforms the Ashtekar canonical one form into

\begin{eqnarray}
\label{CEDE}
\boldsymbol{\theta}_{Ash}=\int_{\Sigma}d^3x\widetilde{\sigma}^i_a\delta{A}^a_i=\int_{\Sigma}d^3x\Psi_{ae}B^i_e\delta{A}^a_i.
\end{eqnarray}

\noindent
We would like to define a configuration variable $X^{ae}$ conjugate to $\Psi_{ae}$ with canonical one form

\begin{eqnarray}
\label{CEDE1}
\boldsymbol{\theta}_{Inst}=\int_{\Sigma}d^3x\Psi_{ae}(x)\delta{X}^{ae}(x).
\end{eqnarray}

\noindent
However, although the variations $\delta{X}^{ae}=B^i_e\delta{A}^a_i$ live in the cotangent space $T^{*}_X(\Gamma_{Inst})$ to the configuration space $\Gamma_{Inst}$, the coordinates $X^{ae}$ do not in general 
exist globally on $\Gamma$ (See e.g. \cite{SOO} and \cite{SOO1}), except for the trace

\begin{eqnarray}
\label{TRACE}
\delta{X}^{11}+\delta{X}^{22}+\delta{X}^{33}=B^i_a\delta{A}^a_i=\delta{I}_{CS}[A],
\end{eqnarray}

\noindent
where $I_{CS}$ is the Chern--Simons functional of the connection $A^a_i$.  Moreover, the symplectic two form in Ashtekar variables is the exact variation of $\boldsymbol{\theta}_{Ash}$

\begin{eqnarray}
\label{DEDED}
\boldsymbol{\omega}_{Ash}=\int_{\Sigma}d^3x{\delta\widetilde{\sigma}^i_a}\wedge{\delta{A}^a_i}=\delta\Bigl(\int_{\Sigma}d^3x\widetilde{\sigma}^i_a\delta{A}^a_i\Bigr)=\delta\boldsymbol{\theta}_{Ash},
\end{eqnarray}

\noindent
whereas in the instanton representation we have

\begin{eqnarray}
\label{DEDED1}
\delta\Bigl(\int_{\Sigma}d^3x\Psi_{ae}B^i_e\delta{A}^a_i\Bigr)
=\int_{\Sigma}d^3x{\delta\Psi_{ae}}\wedge{B^i_e\delta{A}^a_i}+\int_{\Sigma}d^3x\Psi_{ae}{\delta{B}^i_e}\wedge{\delta{A}^a_i}.
\end{eqnarray}

\noindent
Equation (\ref{DEDED1}) is not an exact two form on $\Omega_{Inst}$ unless the second contribution on the right hand side vanishes.  Let us attempt to deduce the allowed configurations of $A^a_i$ for which this may be the case, using the eigenvalues $\lambda_f$ as the fundamental momentum space variables.  There is no loss of generality in taking $\Psi_{ae}$ to be already in diagonal form, hence 

\begin{eqnarray}
\label{ONEFO}
\boldsymbol{\theta}_{Kin}=\sum_a\lambda_{aa}B^i_a\delta{A}^a_i
\end{eqnarray}

\noindent
is the canonical one form at the kinematical level for some $A^a_i$.  The coefficients of $\lambda_{aa}$ in (\ref{ONEFO}) can be split into two terms $B^i_a\delta{A}^a_i=N^a+M^a$ for each $a$, where

\begin{eqnarray}
\label{ONEFO1}
N^a=\sum_{i,j,k}\epsilon^{ijk}(\partial_jA^a_k)\delta{A}^a_i;~~M^a={1 \over 2}\sum_{i,j,k}\epsilon^{ijk}f_{abc}(A^b_jA^c_k)\delta{A}^a_i
\end{eqnarray}

\noindent
with no summation over $a$.  Note that $M^a$ is completely free of spatial gradients of the connection $A^a_i$, while $N^a$ contains spatial gradients.  Dynamical variables containing spatial gradients pose a problem for the full theory, when promoting them to operators for quantization.  Our definition of minisuperspace requires that dynamical variables be spatially homogeneous, depending only on time.  This requires that for a minisuperspace theory, all spatial gradients of the variables must be set to zero.\footnote{Note that this is not the usual definition of minisuperspace via Bianchi groups, which absorb all spatial dependence of the theory into invariant one forms and vector fields.}\par
\indent  
If one could find well-defined configurations in the full theory where all terms containing spatial gradients vanish from the canonical structure even though the spatial gradients are in general nonzero, then one would have the full theory with the advantages of the simplicity of minisuperspace as we have defined it.  Focus first on $N^a$, expanding the individual terms

\begin{eqnarray}
\label{ONEFO2}
N^a=(\partial_2A^a_3)\delta{A}^a_1-(\partial_3A^a_2)\delta{A}^a_1\nonumber\\
+(\partial_3A^a_1)\delta{A}^a_2-(\partial_1A^a_3)\delta{A}^a_2
+(\partial_1A^a_2)\delta{A}^a_3-(\partial_2A^a_1)\delta{A}^a_3.
\end{eqnarray}

\noindent
Now rearrange the terms of (\ref{ONEFO2}) into the form

\begin{eqnarray}
\label{ONEFO3}
N^a=\bigl((\delta{A}^a_3)\partial_1-(\delta{A}^a_1)\partial_3\bigr)A^a_2\nonumber\\
+\bigl((\delta{A}^a_1)\partial_2-(\delta{A}^a_2)\partial_1\bigr)A^a_3+
\bigl((\delta{A}^a_2)\partial_3-(\delta{A}^a_3)\partial_2\bigr)A^a_1.
\end{eqnarray}

\noindent
A moment's reflection of shows that a sufficient condition to make (\ref{ONEFO3}) vanish, is to set two out of three elements of the set $(A^a_1,A^a_2,A^a_3)$ to zero for each $a$.  For 
instance, choosing $A^a_2=A^a_3=0$, which selects two different elements of the 3 by 3 matrix $A^a_i$ from the same row, causes $N_a$ to vanish with no restriction on $A^a_1$.  Performing this for each $a$ leads to the realization that one is free to set six out of the nine elements of $A^a_i$ to zero in the full theory, while still causing $(N_1,N_2,N_3)$ to vanish.  This leaves remaining three nonzero elements $A^a_i$ which is just as well, since there should be three configuration space physical degrees of freedom canonically conjugate to $(\lambda_{11},\lambda_{22},\lambda_{33})$, in order to have a cotangent bundle structure at the kinematical level.  This provides the sought after principle for selecting the configuration space variables needed for quantization.\par
\indent
The question then arises as to which three elements $A^a_i$ to select for the kinematic configuration space $\Gamma_{Kin}$.  If one selects the three $A^a_i$ such that no two elements come from the same 
row $a$ or from the same column $i$ then one has that $\hbox{det}(A^a_i)\neq{0}$, namely that the connection is nondegenerate as a three by three matrix.  Hence we have 
that $N^a=0$ and only $M^a$ contributes to (\ref{ONEFO}).  This is given by

\begin{eqnarray}
\label{ONEFO4}
M^a=(\hbox{det}A)(A^{-1})^i_a\delta{A}^a_i,
\end{eqnarray}

\noindent
where we have used the fact that the structure constants $f_{abc}$ for the Ashtekar variables are numerically the same as the Cartesian epsilon symbol $\epsilon_{abc}$ in writing the determinant.  The canonical one form then is given by

\begin{eqnarray}
\label{ONEFO5}
\boldsymbol{\theta}_{Kin}=\int_{\Sigma}d^3x\lambda_{aa}(\hbox{det}A)(A^{-1})^i_a\delta{A}^a_i
=\int_{\Sigma}d^3x\widetilde{\lambda}_a(A^{-1})^i_a\delta{A}^a_i
\end{eqnarray}

\noindent
where $\widetilde{\lambda}_a=\lambda_{aa}(\hbox{det}A)$ are the densitized version of the eigenvalues of $\Psi_{(ae)}$.  Since $A^a_i$ contains three D.O.F., then the configuration space term can be written in the form

\begin{eqnarray}
\label{ONEFO6}
(A^{-1})^i_a\delta{A}^a_i={{\delta{A}^a_i} \over {A^a_i}}=\delta(\hbox{ln}A^a_i/a_0)=\delta{X}^a_i
\end{eqnarray}

\noindent
where a different choice of $i$ must be made for each $a$.\footnote{The quantity $a_0$ is a numerical constant of mass dimension $[a_0]=1$, needed to make the argument of the logarithm dimensionless.}  The variables $X^a_i$ are globally holonomic, hence they form a good set of coordinates on $\Gamma_{Kin}$ with symplectic two form

\begin{eqnarray}
\label{TWOFO}
\boldsymbol{\omega}_{Kin}=\int_{\Sigma}d^3x{\delta\widetilde{\lambda}_a}\wedge(A^{-1})^i_a\delta{A}^a_i,
\end{eqnarray}

\noindent
where $\boldsymbol{\omega}_{Kin}=\delta\boldsymbol{\theta}_{Kin}$.  These variables are canonically conjugate to the densitized eigenvalues $\lambda_a$, and serve as a basis for quantization of the full theory for configurations where the CDJ matrix is diagonalizable.  In this case one can always perform a $SO(3,C)$ rotation into the diagonal configuration by using the polar decomposition of $\Psi_{ae}$, not including the angles $\vec{\theta}$ as part of the canonical structure.\footnote{In Paper 
IV it is proven that the $SO(3,C)$ angles $\vec{\theta}$ are indeed ignorable in the canonical and in the symplectic structures of the instanton representation.}  Hence the condition $N_a=0$ makes it possible to globally define coordinates corresponding to $M_a$ for the full theory, in direct analogy to minisuperspace.\par
\indent
The quantizable configurations on the kinematic phase space of the instanton representation then imply the following restriction of the configuration space of the Ashtekar variables  

\begin{eqnarray}
\label{BUNDLE15}
A^a_i=I_{f_1f_2f_3}I_{j_1j_2j_3}\Bigl(\delta_{af_1}\delta_{ij_1}A^{f_1}_{j_1}+\delta_{af_2}\delta_{ij_2}A^{f_2}_{j_2}+\delta_{af_3}\delta_{ij_3}A^{f_3}_{j_3}\Bigr),
\end{eqnarray}

\noindent
where $I_{ijk}=1$ for $i\neq{j}\neq{k}$, and zero otherwise.  The actual variables which will be quantized are obtained by functional antidifferentiation of (\ref{ONEFO6}), which yields

\begin{eqnarray}
\label{BUNDLE17}
X^f
=I_{f_1f_2f_3}I_{j_1j_2j_3}\Bigl(\delta_{ff_1}\delta_{ij_1}\hbox{ln}\Bigl({{{A}^{f_1}_{j_1}} \over {a_0}}\Bigr)
+\delta_{ff_2}\delta_{ij_2}A^{f_2}_{j_2}\hbox{ln}\Bigl({{{A}^{f_2}_{j_2}} \over {a_0}}\Bigr)
+\delta_{ff_3}\delta_{ij_3}A^{f_3}_{j_3}\hbox{ln}\Bigl({{{A}^{f_3}_{j_3}} \over {a_0}}\Bigr)\Bigr)
\end{eqnarray}

\noindent
for $f=1,2,3$.  The ranges of the coordinates are $-\infty<\vert{X}^f\vert<\infty$, corresponding to $0<\vert{A}^a_i\vert<\infty$, which guarantee nondegeneracy of $A^a_i$.  The result is that the instanton representation admits a quantization of the full theory on the kinematic phase space we should have canonical commutation relations

\begin{eqnarray}
\label{WESHOULDHAVE}
\bigl[X^f(x,t),\widetilde{\lambda}_g(y,t)\bigr]=\delta^f_g\delta^{(3)}(x,y).
\end{eqnarray}

\noindent
We have shown that on the quantizable configurations the spatial gradients cancel out of the canonical one form $\boldsymbol{\theta}_{Kin}$ in the full theory.  It so happens that precisely on these configurations, the spatial gradients also vanish from the Chern--Simons functional

\begin{eqnarray}
\label{CHERM}
L_{CS}={A^a}\wedge{dA^a}+{2 \over 3}{A}\wedge{A}\wedge{A}\longrightarrow{2 \over 3}{A}\wedge{A}\wedge{A},
\end{eqnarray}

\noindent
which can be shown by a similar argument as above using (\ref{ONEFO2}) and (\ref{ONEFO3}) with $\delta{A}^a_i$ replaced by $A^a_i$.  The result is that when one densitizes $\Psi_{ae}$ by $(\hbox{det}A)$, one is in fact 
densitizing $\Psi_{ae}$ by the Chern--Simons Lagrangian evaluated on the quantizable configurations.

\newpage

\section{Quantization in the instrinsic $SO(3,C)$ frame}

\noindent
The intrinsic $SO(3,C)$ frame is defined as the frame of reference in which the symmetric part of the CDJ matrix $\Psi_{(ae)}$ is diagonalized, and can be associated to the kinematic level of the instanton representation.  We will provide additional arguments that it is possible to carry out a quantization of the full theory with respect to this particular frame.  Let us start from the CDJ Ansatz

\begin{eqnarray}
\label{NOISE}
\widetilde{\sigma}^i_a=\Psi_{ae}B^i_e.
\end{eqnarray}

\noindent
Next, re-write (\ref{NOISE}) using the polar decomposition of $\Psi_{ae}$ 

\begin{eqnarray}
\label{NOISE1}
\widetilde{\sigma}^i_a=(e^{\theta\cdot{T}})_{af}\lambda_f(e^{-\theta\cdot{T}})_{fe}B^i_e+\epsilon_{aed}B^i_e\psi_d,
\end{eqnarray}

\noindent
where $\lambda_f$ are the eigenvalues of $\Psi_{(ae)}$.  Next, multiply (\ref{NOISE1}) by $e^{-\theta\cdot{T}}$, which will have the effect of rotating the index $a$  
into the intrinsic $SO(3,C)$ frame

\begin{eqnarray}
\label{NOISE2}
(e^{-\theta\cdot{T}})_{fa}\widetilde{\sigma}^i_a=\lambda_f(e^{-\theta\cdot{T}})_{fe}B^i_e+(e^{-\theta\cdot{T}})_{fa}\epsilon_{aed}B^i_e\psi_d.
\end{eqnarray}

\noindent
Now make the definition

\begin{eqnarray}
\label{NOISE3}
B^i_e=(e^{\theta\cdot{T}})_{eh}b^i_h;~~\widetilde{P}^i_a=(e^{-\theta\cdot{T}})_{ae}\widetilde{\sigma}^i_e,
\end{eqnarray}

\noindent
where $b^i_h$ is the magnetic field for a `reference' connection $a^a_i$ associated with the intrinsic $SO(3,C)$ frame.  Then $B^i_e$ is a gauge-transformed version of $b^i_h$, which can be parametrized by six degrees of freedom.  In the intrinsic $SO(3,C)$ frame we have that

\begin{eqnarray}
\label{NOISE4}
\widetilde{P}^i_f=\lambda_fb^i_f+(e^{-\theta\cdot{T}})_{fa}(e^{-\theta\cdot{T}})_{he}\epsilon_{aed}b^i_h\psi_d.
\end{eqnarray}

\noindent
Now multiply (\ref{NOISE4}) by $(b^{-1})^g_i$ and use the complex orthogonal property 

\begin{eqnarray}
\label{USETHE}
(e^{-\theta\cdot{T}})_{fa}(e^{-\theta\cdot{T}})_{ge}(e^{-\theta\cdot{T}})_{bh}\epsilon_{aeh}=\epsilon_{fbg},
\end{eqnarray}

\noindent
which yields

\begin{eqnarray}
\label{NOISE5}
(b^{-1})^g_i\widetilde{P}^i_f=\delta_{gf}\lambda_f+\epsilon_{fgb}(e^{-\theta\cdot{T}})_{bd}\psi_d.
\end{eqnarray}

\noindent
At this stage one densitizes (\ref{NOISE5}) by multiplying by the determinant of $A^a_i$, which implies the following Schr\"odinger representation

\begin{eqnarray}
\label{NOISE6}
(\hbox{det}A)(b^{-1})^g_i{\delta \over {\delta{a}^f_i}}
=(\hbox{det}A){\delta \over {\delta{X}^{fg}}}\nonumber\\
=\delta_{gf}{\delta \over {\delta{X}^f}}+\epsilon_{fgb}(e^{-\theta\cdot{T}})_{bd}{\delta \over {\delta{X}^{[d]}}},
\end{eqnarray}

\noindent
where $X^{[d]}$ is the variable conjugate to $\psi_d$, which may not be well-defined.  Let us now apply a counting argument of the degrees of freedom.\footnote{The following notation $Dim(p,q)=(a,b)$ signifies that the complex degrees of freedom per point respectively in the momentum space $p$ and the configuration space $q$ are respectively $a$ and $b$.}  At the unconstrained level, starting from the full unconstrained phase space $\Omega_{Inst}$, we have $Dim(\Psi_{ae},A^a_i)=(9,9)$.  The phase space variables can both be written in terms of a polar decomposition\footnote{For the second line of (\ref{POLARDECOMP}) we have adapted for GR the polar representation of $SU(2)$ Yang--Mills gauge fields presented in \cite{POLAR}.  There are two complex orthogonal matrices, $e^{\vec{\theta}\cdot{T}}$ which rotates the internal index, hence a gauge transformation, and the other matrix $U_{fi}[\vec{\theta}]$ which rotates the spatial index.  The latter is parametrized by three angles $\vec{\phi}=(\phi^1,\phi^2,\phi^3)$.  The physical degrees of freedom are encoded in the three diagonal components $a_f$.  Hence there are a total of nine complex degrees of freedom in $A^a_i$.}

\begin{eqnarray}
\label{POLARDECOMP}
\Psi_{ae}=(e^{\theta\cdot{T}})_{af}\lambda_f(e^{-\theta\cdot{T}})_{fe}+\epsilon_{aed}\psi_d;\nonumber\\
A^a_i=(e^{\theta\cdot{T}})_{af}a_fU_{fi}[\vec{\phi}]-{1 \over 2}\epsilon_{abc}(e^{\theta\cdot{T}})_{fb}\partial_i(e^{\theta\cdot{T}})_{fc}.
\end{eqnarray}

\noindent
Implementation of the diffeomorphism constraint 
sets $\psi_d=0$, with no corresponding reduction of the configuration space, yielding $Dim(\Psi_{ae},A^a_i)_{diff}=(6,9)$.  At this point the rotation into the intrinsic $SO(3,C)$ frame absorbs the angles $\vec{\theta}$ into the definition of the variables, which corresponds to a reduction both of configuration space and momentum space to $Dim(\Psi_{ae},A^a_i)_{Kin}=(3,6)$.  In order to have a cotangent bundle structure at this stage, we need to eliminate three D.O.F. from the configuration space.  By setting three elements of $A^a_i$ to zero such that $(\hbox{det}A)\neq{0}$, we obtain the required structure with $Dim(\Psi_{ae},A^a_i)\equiv{Dim}(\lambda_f,a_f)=(3,3)$ and globally holonomic coordinates.  This is tantamount to setting $\vec{\phi}=0$ in (\ref{POLARDECOMP}), and forms the starting point for a quantization of the physical degrees of freedom of the theory.\footnote{Is is 
known from Paper XIV that $X^{[d]}$ in (\ref{NOISE6}) cannot be defined as coordinates on configuation space $\Gamma$, since they are not integrable, and that there are no D.O.F. conjugate to the angles $\vec{\theta}$.  This implies that the $SO(3,C)$ frame is ignorable in the canonical structure of the instanton representation, and suggests that not more than six phase space variables on $\Omega_{Inst}$ are quantizable.}  Note for the diagonal elements $f=g$, that we have

\begin{eqnarray}
\label{NOISE7}
(\hbox{det}A)(b^{-1})^f_i{\delta \over {\delta{a}^f_i}}={\delta \over {\delta{X}^f}},
\end{eqnarray}

\noindent
which is globally holonomic on configuration space.\par
\indent
The intrinsic frame can be achieved directly from the level of the instanton representation action, starting from 

\begin{eqnarray}
\label{NOISE8}
I_{Inst}=\int{dt}\int_{\Sigma}d^3x\Psi_{ae}B^i_e\dot{A}^a_i+A^a_0B^i_eD_i\Psi_{ae}\nonumber\\
-\epsilon_{ijk}N^iB^j_aB^k_e\Psi_{ae}-N(\hbox{det}B)^{1/2}\sqrt{\hbox{det}\Psi}\bigl(\Lambda+\hbox{tr}\Psi^{-1}\bigr).
\end{eqnarray}

\noindent
Integrating by parts and separating $\Psi_{ae}$ into its symmetric and its antisymmetric parts, we have

\begin{eqnarray}
\label{NOISE9}
I_{Inst}=\int{dt}\int_{\Sigma}d^3x\Psi_{(ae)}B^i_eF^a_{0i}-N(\hbox{det}B)^{1/2}\sqrt{\hbox{det}\Psi}\bigl(\Lambda+\hbox{tr}\Psi^{-1}\bigr)\nonumber\\
+\int{dt}\int_{\Sigma}d^3x\bigl(B^i_eF^a_{0i}-\epsilon_{ijk}N^iB^j_aB^k_e\bigr)\Psi_{ae}.
\end{eqnarray}

\noindent
Implementation of the diffeomorphism constraint eliminates the second line of (\ref{NOISE9}) since $\Psi_{[ae]}=0$ from (\ref{DIFF}).  Using the fact that $\Psi_{ae}=\Psi_{(ae)}$ is symmetric in $a$ and $e$, we can 
now write (\ref{NOISE9}) as

\begin{eqnarray}
\label{NOISE10}
I_{Inst}=\int_{\Sigma}d^4x\Bigl({1 \over 8}\Psi_{ae}F^a_{\mu\nu}F^e_{\rho\sigma}\epsilon^{\mu\nu\rho\sigma}-N(\hbox{det}B)^{1/2}\sqrt{\hbox{det}\Psi}\bigl(\Lambda+\hbox{tr}\Psi^{-1}\bigr)\Bigr),
\end{eqnarray}

\noindent
Equation (\ref{NOISE4}) effectively appends the Hamiltonian constraint to an object which resembles a topological ${F}\wedge{F}$ term, with $\Psi_{ae}$ replacing the Cartan--Killing form.  Next we will implement the Gauss' law constraint, using the polar decomposition of the CDJ matrix and the fact that the Hamiltonian constraint 
for symmetric $\Psi_{ae}$ is $SO(3,C)$ invariant.  First, note that the first term of (\ref{NOISE10}) can be written as

\begin{eqnarray}
\label{NOISE11}
{1 \over 8}\int_Md^4x\lambda_f(e^{-\theta\cdot{T}})_{fa}(e^{-\theta\cdot{T}})_{fe}F^a_{\mu\nu}[A]F^e_{\rho\sigma}[A],
\end{eqnarray}

\noindent
such that each curvature is rotated in its internal index.  This rotation corresponds to the $SO(3,C)$ gauge transformation of $F^a_{\mu\nu}[A]$ into a new curvature $f^a_{\mu\nu}[a]$ for some connection $a^a_{\mu}dx^{\mu}$, which is just $A^a=A^a_{\mu}dx^{\mu}$ in another gauge.  The relation is given by

\begin{eqnarray}
\label{NOISE12}
a=(e^{-\theta\cdot{T}})(A+d)(e^{\theta\cdot{T}}).
\end{eqnarray}

\noindent
Hence, the rotation of (\ref{NOISE10}) into this $SO(3,C)$ frame yields

\begin{eqnarray}
\label{NOISE13}
I_{Inst}=\int_Md^4x\Bigl({1 \over 8}\lambda_ff^f_{\mu\nu}[a]f^f_{\rho\sigma}[a]\epsilon^{\mu\nu\rho\sigma}\nonumber\\
-N(\hbox{det}b)^{1/2}\sqrt{\lambda_1\lambda_2\lambda_3}\Bigl(\Lambda+{1 \over {\lambda_1}}+{1 \over {\lambda_2}}+{1 \over {\lambda_3}}\Bigr)\Bigr).
\end{eqnarray}

\noindent
As shown in Paper II, we can implement the Gauss' law constraint (\ref{GAUSSLAW}) and choose the gauge $a_0=0$, which puts (\ref{NOISE13}) into canonical form.  From this point we can implement the Hamiltonian constraint 
and use (\ref{NOISE3}) to construct a Hamilton--Jacobi functional.

\subsection{Canonical equivalence to the Ashtekar variables}

We will now show that the phase space of the instanton representation on globally holonomic configurations is equivalent to the physical phase space.  The commutation relations for the Ashtekar variables are given by

\begin{eqnarray}
\label{THECOMMUTATION}
\bigl[A^a_i(x),\widetilde{\sigma}^j_b(y)\bigr]=\delta^a_b\delta^j_i\delta^{(3)}(x,y),
\end{eqnarray}

\noindent
where we have omitted the time dependence to avoid cluttering up the notation.  Let us now substitute the CDJ Ansatz $\widetilde{\sigma}^i_a=\Psi_{ae}B^i_e$ into (\ref{THECOMMUTATION})

\begin{eqnarray}
\label{THECOMMUTATION1}
\bigl[A^a_i(x),\Psi_{be}(y)B^j_e(y)\bigr]=\delta^a_b\delta^j_i\delta^{(3)}(x,y).
\end{eqnarray}

\noindent
We will now multiply (\ref{THECOMMUTATION1}) by $A^c_j(y)$ in the following form

\begin{eqnarray}
\label{THECOMMUTATION2}
\bigl[A^a_i(x),\Psi_{be}(y)B^j_e(y)A^c_j(y)\bigr]=\delta^a_bA^c_i(y)\delta^{(3)}(x,y),
\end{eqnarray}

\noindent
which is allowed since $[A^a_i,A^c_j]=0$ for the Ashtekar connection.  Define the magnetic helicity density matrix $C_{ce}=A^b_jB^j_e$, written in component form as

\begin{eqnarray}
\label{THECOMMUTATION3}
C_{ce}=\epsilon^{ijk}A^c_i\partial_jA^e_k+\delta_{ce}(\hbox{det}A),
\end{eqnarray}

\noindent
which has a diagonal part free of spatial gradients and an off-diagonal part containing spatial gradients.  Then the commutation relations read

\begin{eqnarray}
\label{THECOMMUTATION4}
\bigl[A^a_i(x),\Psi_{be}(y)C_{ce}(y)\bigr]=\delta^a_bA^c_i(y)\delta^{(3)}(x,y).
\end{eqnarray}

\noindent
The kinematic configuration space $\Gamma_{Kin}$ must have three degrees of freedom per point.\footnote{This is nine total degrees of freedom, minus three corresponding to $G_a$, and minus three corresponding to $H_i$.}  Let us choose, without loss of generality, for these D.O.F. to be the three diagonal elements $A^a_i=\delta^a_iA^a_a$.  Then we can set $a=i$ in (\ref{THECOMMUTATION4}) to obtain

\begin{eqnarray}
\label{THECOMMUTATION5}
\bigl[A^a_a(x),\Psi_{be}(y)C_{ce}(y)\bigr]=\delta^a_bA^c_a(y)\delta^{(3)}(x,y).
\end{eqnarray}

\noindent
Since $A^a_i$ is diagonal by supposition, then the only nontrivial contribution to (\ref{THECOMMUTATION5}) occurs for $a=c$.  Since $a=b$ also is the only nontrivial contribution, it follows that $b=c$ as well.  Hence the commutation relations for diagonal connection are given by

\begin{eqnarray}
\label{THECOMMUTATION6}
\bigl[A^a_a(x),\Psi_{be}(y)C_{be}(y)\bigr]=\delta^a_b\delta{A}^b_b(y)\delta^{(3)}(x,y).
\end{eqnarray}

\noindent
Substituting (\ref{THECOMMUTATION3}) subject to a diagonal connection into (\ref{THECOMMUTATION6}) we have

\begin{eqnarray}
\label{THECOMMUTATION7}
\sum_{e=1}^3\bigl[A^a_a(x),\Psi_{be}(y)\delta_{be}(\hbox{det}A)\bigr]\nonumber\\
+\sum_{e=1}^3\bigl[A^a_a(x),\Psi_{be}(y)\epsilon^{bje}A^b_b\partial_jA^e_e\bigr]=\delta^a_bA^b_b(y)\delta^{(3)}(x,y),
\end{eqnarray}

\noindent
which has split up into two terms.  We have been explicit in putting in the summation symbol to indicate that $e$ is a dummy index, while $a$ and $b$ are not.  There are two cases to consider, $e=b$ and $e\neq{b}$.  For $e\neq{b}$ the first term of (\ref{THECOMMUTATION7}) vanishes, leaving remaining the second term.  Since the right hand side stays the same, then this would correspond to the commutation relations for a CDJ matrix whose diagonal components are zero.  For the second possibility $e=b$ the second term of (\ref{THECOMMUTATION7}) vanishes while the first term survives, with the right hand side the same as before.  This case occurs only if the CDJ matrix $\Psi_{ae}$ is diagonal.  Let us choose $\Psi_{ae}=Diag(\lambda_1,\lambda_2,\lambda_3)$ as the diagonal matrix of eigenvalues, then (\ref{THECOMMUTATION7}) reduces to

\begin{eqnarray}
\label{THECOMMUTATION8}
\bigl[A^a_a(x),\lambda_b(y)(\hbox{det}A(y))\bigr]=\delta^a_bA^a_a(y)\delta^{(3)}(x,y).
\end{eqnarray}

\noindent
The conclusion is that in order for (\ref{THECOMMUTATION8}) to have arisen from (\ref{THECOMMUTATION}), that: (i) The antisymmetric part of $\Psi_{ae}$ must be zero, namely, the diffeomorphism constraint must be satisfied. (ii) The symmetric off-diagonal part of $\Psi_{ae}$ is not part of the commutation relations on the diffeomorphism invariant phase space $\Omega_{diff}$.  Given the eigenvalues $\lambda_f$ on this space, the Gauss' law constraint can be solved separately from the quantization process.  The choice of diagonal $A^a_a$ is consistent with the implementation of the kinematic constraints, which means that only the Hamiltonian constraint is necessary to obtain the physical phase space $\Omega_{Phys}$.\par
\indent
Equation (\ref{THECOMMUTATION8}) is not canonical owing to the field-dependence on the right hand side,\footnote{While (\ref{THECOMMUTATION8}) are not canonical commutation relations, they are affine commutation relations which serve as an intermediate step in the formulation of canonical commutation relations.  Affine commutation relations have been used by Klauder in \cite{KLAUDER} in the affine quantum gravity programme, and are viable as well in the instanton representation.} however it implies canonical relations according to the following

\begin{eqnarray}
\label{THECOMMUTATION9}
\bigl[\hbox{ln}\Bigl({{A^a_a(x)} \over {a_0}}\Bigr),\Pi_b(y)\Bigr]=(A^{-1}(x))^a_a\bigl[A^a_a(x),\Pi_b(y)\bigr]=(A^{-1}(y))^a_a\bigl[A^a_a(x),\Pi_b(y)\bigr],
\end{eqnarray}

\noindent
where we have defined $\Pi_b=\lambda_b(\hbox{det}A)$.  The first step of (\ref{THECOMMUTATION9}) follows from the chain rule, and the second step follows from the fact that the only nontrivial contribution comes from $x=y$.  Comparison of (\ref{THECOMMUTATION9}) with (\ref{THECOMMUTATION8}) implies that the canonical version of (\ref{THECOMMUTATION8}) is given by

\begin{eqnarray}
\label{THECOMMUTATION10}
\bigl[X^a(x),\Pi_b(y)\bigr]=\delta^a_b\delta^{(3)}(x,y),
\end{eqnarray}

\noindent
where we have defined $X^a=\hbox{ln}(A^a_a/a_0)$.\footnote{This is because the inverse is the same as the reciprocal for a diagonal connection.  The coordinate ranges are $\infty<{X}<\infty$, which corresponds to $0<A^f_f<\infty$, which is a subset of the latter.  To utilize the full range of $A^a_i$, which includes the degenerate cases, one may instead use (\ref{THECOMMUTATION8}).}  We have shown that $\Omega_{Kin}$ of the instanton representation admits a cotangent bundle structure with diagonal connection $A^a_a(x)$.  It happens from (\ref{THECOMMUTATION}) that $A^a_a(x)$ is canonically conjugate to $\widetilde{\sigma}^a_a(x)$.  Since the instanton representation maps to the Ashtekar formalism and vice versa on the unreduced phase space for nondegenerate $B^i_a$, it 
follows that (\ref{THECOMMUTATION10}) corresponds as well to the kinematic phase space of the Ashtekar variables for $(\hbox{det}A)\neq{0}$, six total degrees of freedom per point, where the variables are diagonal.  The bonus is that all the kinematic constraints have been implemented, leaving behind the Hamiltonian constraint which in the instanton representation is easy to solve.\par  
\indent
We have previously shown that each nondegnerate $A^a_i$ with six out of nine elements set to zero admits globally holonomic coordinates in the instanton representation.  Since $A^a_i$ serves also as the configuration variable for the Ashtekar phase space $\Omega_{Ash}$, it follows that on this subspace the densitized triad must also be nondegenerate.  Hence

\begin{eqnarray}
\label{THCOM}\bigl[A^f_f(x,t),\widetilde{\sigma}^g_g(y,t)\bigr]=\delta^f_g\delta^{(3)}(x,y).
\end{eqnarray}

\noindent
The result is that the kinematic phase space of the instanton representation corresponds to nondegenerate triads, in the original Ashtekar variables, at the level prior to implementation of the Hamiltonian constraint.

\newpage

\section{Verification of the quantizable configurations for the instanton representation}

\noindent
We will now demonstrate the results of the previous subsection by explicitly computing the allowed configurations which may be used globally as configuration space variables for 
quantization of the instanton representation.  This will lead us to six quantizable configurations in the full theory.  Let us define ${a}^a_i\equiv{A}^a_i$ as the resulting spatial connection, in terms of which we will derive the canonical structure.  It is convenient for bookkeeping purposes, starting from the Ashtekar magnetic field 

\begin{eqnarray}
\label{MAGNETICFIELD}
B^i_a=\epsilon^{ijk}\partial_jA^a_k+{1 \over 2}\epsilon^{ijk}f_{abc}A^b_jA^c_k,
\end{eqnarray}

\noindent
to write out explicitly the individual components and group into

\begin{eqnarray}
\label{MAGNET}
B^1_1=\partial_2A^1_3-\partial_3A^1_2+A^2_2A^3_3-A^2_3A^3_2;\nonumber\\
B^2_2=\partial_3A^2_1-\partial_1A^2_3+A^3_3A^1_1-A^3_1A^1_3;\nonumber\\
B^3_3=\partial_1A^3_2-\partial_2A^3_1+A^1_1A^2_2-A^1_2A^2_1
\end{eqnarray}

\noindent
for the diagonal components, and

\begin{eqnarray}
\label{MAGNET1}
B^1_2=\partial_2A^2_3-\partial_3A^2_2+A^3_2A^1_3-A^3_3A^1_2;\nonumber\\
B^2_3=\partial_3A^3_1-\partial_1A^3_3+A^1_3A^2_1-A^1_1A^2_3;\nonumber\\
B^3_1=\partial_1A^1_2-\partial_2A^1_1+A^2_1A^3_2-A^2_2A^3_1
\end{eqnarray}

\noindent
and

\begin{eqnarray}
\label{MAGNET2}
B^2_1=\partial_3A^1_1-\partial_1A^1_3+A^2_3A^3_1-A^2_1A^3_3;\nonumber\\
B^3_2=\partial_1A^2_2-\partial_2A^2_1+A^3_1A^1_2-A^3_2A^1_1;\nonumber\\
B^1_3=\partial_2A^3_3-\partial_3A^3_2+A^1_2A^2_3-A^1_3A^2_2
\end{eqnarray}

\noindent
for the off-diagonal components.  Using (\ref{MAGNET}), (\ref{MAGNET1}) and (\ref{MAGNET2}) we will explicitly determine the configurations that yield the desired canonical structure.\par
\indent
The first configuration, where $A^a_i=\delta_{ai}A^a_i$ is diagonal, is given by

\begin{displaymath}
A^a_i=
\left(\begin{array}{ccc}
A^1_1 & 0 & 0\\
0 & A^2_2 & 0\\
0 & 0 & A^3_3\\
\end{array}\right);~~
B^i_a=
\left(\begin{array}{ccc}
A^2_2A^3_3 & -\partial_3A^2_2 & \partial_2A^3_3\\
\partial_3A^1_1 & A^3_3A^1_1 & -\partial_1A^3_3\\
-\partial_2A^1_1 & \partial_1A^2_2 & A^1_1A^2_2\\
\end{array}\right);
\end{displaymath}

\begin{displaymath}
B^i_e\dot{A}^a_i=
\left(\begin{array}{ccc}
A^2_2A^3_3\dot{A}^1_1 & -(\partial_3A^2_2)\dot{A}^2_2 & (\partial_2A^3_3)\dot{A}^3_3\\
(\partial_3A^1_1)\dot{A}^1_1 & A^3_3A^1_1\dot{A}^2_2 & -(\partial_1A^3_3)\dot{A}^3_3\\
-(\partial_2A^1_1)\dot{A}^1_1 & (\partial_1A^2_2)\dot{A}^2_2 & A^1_1A^2_2\dot{A}^3_3.
\end{array}\right)
\end{displaymath}

\noindent
Upon contraction with a diagonal CDJ matrix $\Psi_{ae}=\delta_{ae}\lambda_{ee}$ this leads to the canonical structure

\begin{eqnarray}
\label{CONTRACT1}
\lambda_{11}A^2_2A^3_3\dot{A}^1_1+\lambda_{22}A^3_3A^1_1\dot{A}^2_2+\lambda_{33}A^1_1A^2_2\dot{A}^3_3\nonumber\\
=(A^1_1A^2_2A^3_3)\Bigl[\lambda_{11}\Bigl({{\dot{A}^1_1} \over {A^1_1}}\Bigr)+\lambda_{22}\Bigl({{\dot{A}^2_2} \over {A^2_2}}\Bigr)+\lambda_{33}\Bigl({{\dot{A}^3_3} \over {A^3_3}}\Bigr)\Bigr],
\end{eqnarray}

\noindent
where $\hbox{det}A=A^1_1A^2_2A^3_3$.  Note in (\ref{CONTRACT1}) that all spatial gradients of $A^a_i$ have been cancelled out.  Making the definitions

\begin{eqnarray}
\label{MAKINGTHE}
\Pi_f=\lambda_{ff}(\hbox{det}A);~~X^1=\hbox{ln}\Bigl({{A^1_1} \over {a_0}}\Bigr);~~X^2=\hbox{ln}\Bigl({{A^2_2} \over {a_0}}\Bigr);~~
X^3=\hbox{ln}\Bigl({{A^3_3} \over {a_0}}\Bigr),
\end{eqnarray}

\noindent
this yields the symplectic two form

\begin{eqnarray}
\label{SYMPLECTICTWO}
\boldsymbol{\Omega}_{Kin}=\int_{\Sigma}{\delta\Pi_f(x)}\wedge{\delta{X}^f(x)}=\delta\Bigl(\int_{\Sigma}\Pi_f(x)\delta{X}^f(x)\Bigr)=\delta\boldsymbol{\theta}_{Kin},
\end{eqnarray}

\noindent
which is the exact functional variation of the canonical oneform $\boldsymbol{\theta}_{Kin}$ on the kinematic phase space.  This implies canonical commutation relations 

\begin{eqnarray}
\label{DEFINITION}
\bigl[X^f(x,t),\Pi_f(y,t)\bigr]=\delta^f_g\delta^{(3)}(x,y).
\end{eqnarray}

\noindent
Hence we have obtained globally holonomic coordinates on the kinematic phase space $\Omega_{Kin}$ of the instanton representation, even though such coordinates do not exist on the full phase 
space $\Omega$.  Equation (\ref{MAKINGTHE}) provides the degrees of freedom which can be used for quantization of the full theory, and ranges of the coordinates are

\begin{eqnarray}
\label{THERANGES}
-\infty<\vert{X}^f\vert<\infty;~~0<\vert{A}^f_f\vert<\infty.
\end{eqnarray}

\noindent
Note that the canonical commutation relations (\ref{DEFINITION}) can also be written in the form

\begin{eqnarray}
\label{COINS}
\bigl[A^f_f(x,t),\Pi_g(y,t)\bigr]=\delta^f_gA^g_g\delta^{(3)}(x,y),
\end{eqnarray}

\noindent
which are affine commutation relations analogous to the type introduced in \cite{KLAUDER}.  We will refer to such configurations, where $A^a_i$ is nondegenerate and has three nonzero entries, as quantizable.  In (\ref{COINS}) we have treated $\Pi_f$, the densitized eigenvalues, as the fundamental momentum space variable.  But it is really a composite variable, and (\ref{COINS}) can be written in self-adjoint form as

\begin{eqnarray}
\label{RIGOUROUS}
\bigl[\hat{A}^f_f(x,t),{1 \over 2}\bigl(\hat{\lambda}_g(y,t)\hat{(\hbox{det}A)}+\hat{(\hbox{det}A)}\hat{\lambda}_g(y,t)\bigr)\bigr]=\delta^f_gA^g_g\delta^{(3)}(x,y).
\end{eqnarray}

\noindent
In the original Ashtekar variables this corresponds to canonical commutation relations

\begin{displaymath}
\Biggl[
\left(\begin{array}{ccc}
A^1_1(x,t) & 0 & 0\\
0 & A^2_2(x,t) & 0\\
0 & 0 & A^3_3(x,t)\\
\end{array}\right),
\left(\begin{array}{ccc}
\widetilde{\sigma}^1_1(y,t) & 0 & 0\\
0 & \widetilde{\sigma}^2_2(y,t) & 0\\
0 & 0 & \widetilde{\sigma}^3_3(z,t)\\
\end{array}\right)\Biggr]
=\left(\begin{array}{ccc}
1 & 0 & 0\\
0 &1 & 0\\
0 & 0 & 1\\
\end{array}\right)
\delta^{(3)}(x,y),
\end{displaymath}

\noindent
which involves only the corresponding nondegenerate components of the densitized triad on the kinematical phase space.  The conclusion is that upon implementation of the kinematic constraints the instanton representation can be quantized in the full theory, and maps to the corresponding quantization on the Ashtekar phase space evaluated on nondegenerate triads.

\subsection{Second quantizable configuration}

The second case is given by

\begin{displaymath}
A^a_i=
\left(\begin{array}{ccc}
A^1_1 & 0 & 0\\
0 & 0 & A^2_3\\
0 & A^3_2 & 0\\
\end{array}\right);~~
B^i_a=
\left(\begin{array}{ccc}
-A^2_3A^3_2 & \partial_2A^2_3 & -\partial_3A^3_2\\
\partial_3A^1_1 & -\partial_1A^2_3 & -A^1_1A^2_3\\
-\partial_2A^1_1 & -A^3_2A^1_1 & \partial_1A^3_2\\
\end{array}\right);
\end{displaymath}

\begin{displaymath}
B^i_e\dot{A}^a_i=
\left(\begin{array}{ccc}
-A^2_3A^3_2\dot{A}^1_1 & -(\partial_3A^3_2)\dot{A}^3_2 & (\partial_2A^2_3)\dot{A}^2_3\\
(\partial_3A^1_1)\dot{A}^1_1 & -A^1_1A^2_3\dot{A}^3_2 & -(\partial_1A^2_3)\dot{A}^2_3\\
-(\partial_2A^1_1)\dot{A}^1_1 & (\partial_1A^3_2)\dot{A}^3_2 & -A^3_2A^1_1\dot{A}^2_3\\
\end{array}\right)
.
\end{displaymath}

\noindent
Upon contraction with a diagonal CDJ matrix $\Psi_{ae}=\delta_{ae}\lambda_e$ this leads to the canonical structure

\begin{eqnarray}
\label{CONTRACT2}
-\lambda_{11}A^2_3A^3_2\dot{A}^1_1-\lambda_{22}A^1_1A^2_3\dot{A}^3_2-\lambda_{33}A^3_2A^1_1\dot{A}^2_3\nonumber\\
=-(A^2_3A^3_2A^1_1)\Bigl[\lambda_{11}\Bigl({{\dot{A}^1_1} \over {A^1_1}}\Bigr)+\lambda_{22}\Bigl({{\dot{A}^3_2} \over {A^3_2}}\Bigr)+\lambda_{33}\Bigl({{\dot{A}^2_3} \over {A^2_3}}\Bigr)\Bigr].
\end{eqnarray}

\noindent
where $\hbox{det}A=A^2_3A^3_2A^1_1$.  Making the definitions

\begin{eqnarray}
\label{MAKINGTHE2}
\Pi_f=\lambda_{ff}(\hbox{det}A);~~X^1=\hbox{ln}\Bigl({{A^1_1} \over {a_0}}\Bigr);~~X^2=\hbox{ln}\Bigl({{A^2_3} \over {a_0}}\Bigr);~~
X^3=\hbox{ln}\Bigl({{A^3_2} \over {a_0}}\Bigr),
\end{eqnarray}

\noindent
this yields the canonical commutation relations 

\begin{eqnarray}
\label{DEFINITION2}
\bigl[X^f(x,t),\Pi_f(y,t)\bigr]=\delta^f_g\delta^{(3)}(x,y).
\end{eqnarray}

\noindent
Hence we have obtained globally holonomic coordinates on the kinematic phase space $\Omega_{Kin}$ of the instanton representation, even though such coordinates do not exist on the full phase 
space $\Omega$.  Equation (\ref{MAKINGTHE2}) provides the degrees of freedom which can be used for quantization of the full theory.  This can also be written in the form

\begin{eqnarray}
\label{COINS2}
\bigl[A^f_f(x,t),\Pi_g(y,t)\bigr]=\delta^f_gA^g_g\delta^{(3)}(x,y),
\end{eqnarray}

\noindent
which are affine commutation relations.  We will refer to such configurations, where $A^a_i$ is nondegenerate and has three nonzero entries, as quantizable.  In the original Ashtekar variables this corresponds to canonical commutation relations

\begin{displaymath}
\Biggl[
\left(\begin{array}{ccc}
A^1_1(x,t) & 0 & 0\\
0 & 0 & A^2_3(x,t)\\
0 & A^3_2(x,t) & 0\\
\end{array}\right),
\left(\begin{array}{ccc}
\widetilde{\sigma}^1_1(y,t) & 0 & 0\\
0 & 0 & \widetilde{\sigma}^2_3(y,t)\\
0 & \widetilde{\sigma}^3_2(y,t) & 0\\
\end{array}\right)\Biggr]
=\left(\begin{array}{ccc}
1 & 0 & 0\\
0 &0 & 1\\
0 & 1 & 0\\
\end{array}\right)
\delta^{(3)}(x,y),
\end{displaymath}

\subsection{Third quantizable configuration}

The third case is given by

\begin{displaymath}
A^a_i=
\left(\begin{array}{ccc}
0 & 0 & A^1_3\\
0 & A^2_2 & 0\\
A^3_1 & 0 & 0\\
\end{array}\right);~~
B^i_a=
\left(\begin{array}{ccc}
\partial_2A^1_3 & -\partial_3A^2_2 & -A^1_3A^2_2\\
-\partial_1A^1_3 & -A^3_1A^1_3 & -\partial_3A^3_1\\
-A^2_2A^3_1 & \partial_1A^2_2 & -\partial_2A^3_1\\
\end{array}\right);
\end{displaymath}

\begin{displaymath}
B^i_e\dot{A}^a_i=
\left(\begin{array}{ccc}
-A^1_3A^2_2\dot{A}^3_1 & -(\partial_3A^2_2)\dot{A}^2_2 & (\partial_2A^1_3)\dot{A}^1_3\\
(\partial_3A^3_1)\dot{A}^3_1 & -A^3_1A^1_3\dot{A}^2_2 & -(\partial_1A^1_3)\dot{A}^1_3\\
-(\partial_2A^3_1)\dot{A}^3_1 & (\partial_1A^2_2)\dot{A}^2_2 & -A^2_2A^3_1\dot{A}^1_3.
\end{array}\right)
\end{displaymath}

\noindent
Upon contraction with a diagonal CDJ matrix $\Psi_{ae}=\delta_{ae}\lambda_{ee}$ this leads to the canonical structure

\begin{eqnarray}
\label{CONTRACT3}
-\lambda_{11}A^1_3A^2_2\dot{A}^3_1-\lambda_{22}A^3_1A^1_3\dot{A}^2_2-\lambda_{33}A^2_2A^3_1\dot{A}^1_3\nonumber\\
=-(A^1_3A^2_2A^3_1)\Bigl[\lambda_{11}\Bigl({{\dot{A}^3_1} \over {A^3_1}}\Bigr)+\lambda_{22}\Bigl({{\dot{A}^2_2} \over {A^2_2}}\Bigr)+\lambda_{33}\Bigl({{\dot{A}^1_3} \over {A^1_3}}\Bigr)\Bigr],
\end{eqnarray}

\noindent
where $\hbox{det}A=A^1_3A^2_2A^3_1$.  Making the definitions

\begin{eqnarray}
\label{MAKINGTHE3}
\Pi_f=\lambda_{ff}(\hbox{det}A);~~X^1=\hbox{ln}\Bigl({{A^3_1} \over {a_0}}\Bigr);~~X^2=\hbox{ln}\Bigl({{A^2_2} \over {a_0}}\Bigr);~~
X^3=\hbox{ln}\Bigl({{A^1_3} \over {a_0}}\Bigr),
\end{eqnarray}

\noindent
this yields the canonical commutation relations 

\begin{eqnarray}
\label{DEFINITION3}
\bigl[X^f(x,t),\Pi_f(y,t)\bigr]=\delta^f_g\delta^{(3)}(x,y).
\end{eqnarray}

\noindent
Hence we have obtained globally holonomic coordinates on the kinematic phase space $\Omega_{Kin}$ of the instanton representation, even though such coordinates do not exist on the full phase 
space $\Omega$.  Equation (\ref{MAKINGTHE3}) provides the degrees of freedom which can be used for quantization of the full theory.  This can also be written in the form

\begin{eqnarray}
\label{COINS3}
\bigl[A^f_f(x,t),\Pi_g(y,t)\bigr]=\delta^f_gA^g_g\delta^{(3)}(x,y),
\end{eqnarray}

\noindent
which are affine commutation relations.  We will refer to such configurations, where $A^a_i$ is nondegenerate and has three nonzero entries, as quantizable.  In the original Ashtekar variables this corresponds to canonical commutation relations

\begin{displaymath}
\Biggl[
\left(\begin{array}{ccc}
0 & 0 & A^1_3(x,t)\\
0 & A^2_2(x,t) & 0\\
A^3_1(x,t) & 0 & 0\\
\end{array}\right),
\left(\begin{array}{ccc}
0 & 0 & \widetilde{\sigma}^1_3(y,t)\\
0 & \widetilde{\sigma}^2_2(y,t) & 0\\
\widetilde{\sigma}^3_1(y,t) & 0 & 0\\
\end{array}\right)\Biggr]
=\left(\begin{array}{ccc}
0 & 0 & 1\\
0 & 1 & 0\\
1 & 0 & 0\\
\end{array}\right)
\delta^{(3)}(x,y),
\end{displaymath}

\subsection{Fourth quantizable configuration}

The fourth case, which concludes the configurations containing at least one diagonal connection element, is given by

\begin{displaymath}
A^a_i=
\left(\begin{array}{ccc}
0 & A^1_2 & 0\\
A^2_1& 0 & 0\\
0 & 0 & A^3_3\\
\end{array}\right);~~
B^i_a=
\left(\begin{array}{ccc}
-\partial_3A^1_2 & -A^3_3A^1_2 & \partial_2A^3_3\\
-A^2_1A^3_3 & \partial_3A^2_1 & -\partial_1A^3_3\\
\partial_1A^1_2 & -\partial_2A^2_1 & -A^1_2A^2_1\\
\end{array}\right);
\end{displaymath}

\begin{displaymath}
B^i_e\dot{A}^a_i=
\left(\begin{array}{ccc}
-A^3_3A^1_2\dot{A}^2_1 & -(\partial_3A^1_2)\dot{A}^1_2 & (\partial_2A^3_3)\dot{A}^3_3\\
(\partial_3A^2_1)\dot{A}^2_1 & -A^2_1A^3_3\dot{A}^1_2 & -(\partial_1A^3_3)\dot{A}^3_3\\
-(\partial_2A^2_1)\dot{A}^2_1 & (\partial_1A^1_2)\dot{A}^1_2 & -A^1_2A^2_1\dot{A}^3_3
\end{array}\right)
.
\end{displaymath}

\noindent
Upon contraction with a diagonal CDJ matrix $\Psi_{ae}=\delta_{ae}\lambda_{ee}$ this leads to the canonical structure

\begin{eqnarray}
\label{CONTRACT4}
-\lambda_{11}A^3_3A^1_2\dot{A}^2_1-\lambda_{22}A^2_1A^3_3\dot{A}^1_2-\lambda_{33}A^1_2A^2_1\dot{A}^3_3\nonumber\\
=-(A^3_3A^1_2A^2_1)\Bigl[\lambda_{11}\Bigl({{\dot{A}^2_1} \over {A^2_1}}\Bigr)+\lambda_{22}\Bigl({{\dot{A}^1_2} \over {A^1_2}}\Bigr)+\lambda_{33}\Bigl({{\dot{A}^3_3} \over {A^3_3}}\Bigr)\Bigr].
\end{eqnarray}

\noindent
where $\hbox{det}A=A^3_3A^1_2A^2_1$.  Making the definitions

\begin{eqnarray}
\label{MAKINGTHE4}
\Pi_f=\lambda_{ff}(\hbox{det}A);~~X^1=\hbox{ln}\Bigl({{A^3_1} \over {a_0}}\Bigr);~~X^2=\hbox{ln}\Bigl({{A^2_2} \over {a_0}}\Bigr);~~
X^3=\hbox{ln}\Bigl({{A^1_3} \over {a_0}}\Bigr),
\end{eqnarray}

\noindent
this yields the canonical commutation relations 

\begin{eqnarray}
\label{DEFINITION4}
\bigl[X^f(x,t),\Pi_f(y,t)\bigr]=\delta^f_g\delta^{(3)}(x,y).
\end{eqnarray}

\noindent
Hence we have obtained globally holonomic coordinates on the kinematic phase space $\Omega_{Kin}$ of the instanton representation, even though such coordinates do not exist on the full phase 
space $\Omega$.  Equation (\ref{MAKINGTHE4}) provides the degrees of freedom which can be used for quantization of the full theory.  This can also be written in the form

\begin{eqnarray}
\label{COINS4}
\bigl[A^f_f(x,t),\Pi_g(y,t)\bigr]=\delta^f_gA^g_g\delta^{(3)}(x,y),
\end{eqnarray}

\noindent
which are affine commutation relations.  We will refer to such configurations, where $A^a_i$ is nondegenerate and has three nonzero entries, as quantizable.  In the original Ashtekar variables this corresponds to canonical commutation relations

\begin{displaymath}
\Biggl[
\left(\begin{array}{ccc}
0 & A^1_2(x,t) & 0\\
A^2_1(x,t) & 0 & 0\\
0 & 0 & A^3_3(x,t)\\
\end{array}\right),
\left(\begin{array}{ccc}
0 & \widetilde{\sigma}^1_2(y,t) & 0\\
\widetilde{\sigma}^2_1(y,t) & 0 & 0\\
0 & 0 & \widetilde{\sigma}^3_3(y,t)\\
\end{array}\right)\Biggr]
=\left(\begin{array}{ccc}
0 & 1 & 0\\
1 &0 & 0\\
0 & 0 & 1\\
\end{array}\right)
\delta^{(3)}(x,y),
\end{displaymath}

\subsection{Fifth quantizable configuration}

The fifth case, involving the even permutations, is given by

\begin{displaymath}
A^a_i=
\left(\begin{array}{ccc}
0 & A^1_2 & 0\\
0 & 0 & A^2_3\\
A^3_1 & 0 & 0\\
\end{array}\right);~~
B^i_a=
\left(\begin{array}{ccc}
-\partial_3A^1_2 & \partial_2A^2_3 & A^1_2A^2_3\\
A^2_3A^3_1 & -\partial_1A^2_3 & \partial_3A^3_1\\
\partial_1A^1_2 & A^3_1A^1_2 & -\partial_2A^3_1\\
\end{array}\right);
\end{displaymath}

\begin{displaymath}
B^i_e\dot{A}^a_i=
\left(\begin{array}{ccc}
A^1_2A^2_3\dot{A}^3_1 & -(\partial_3A^1_2)\dot{A}^1_2 & (\partial_2A^2_3)\dot{A}^2_3\\
(\partial_3A^3_1)\dot{A}^3_1 & A^2_3A^3_1\dot{A}^1_2 & -(\partial_1A^2_3)\dot{A}^2_3\\
-(\partial_2A^3_1)\dot{A}^3_1 & (\partial_1A^1_2)\dot{A}^1_2 & A^3_1A^1_2\dot{A}^2_3
\end{array}\right)
.
\end{displaymath}

\noindent
Upon contraction with a diagonal CDJ matrix $\Psi_{ae}=\delta_{ae}\lambda_{ee}$ this leads to the canonical structure

\begin{eqnarray}
\label{CONTRACT4}
\lambda_{11}A^1_2A^2_3\dot{A}^3_1+\lambda_{22}A^2_3A^3_1\dot{A}^1_2+\lambda_{33}A^3_1A^1_2\dot{A}^2_3\nonumber\\
=(A^1_2A^2_3A^3_1)\Bigl[\lambda_{11}\Bigl({{\dot{A}^3_1} \over {A^3_1}}\Bigr)+\lambda_{22}\Bigl({{\dot{A}^1_2} \over {A^1_2}}\Bigr)+\lambda_{33}\Bigl({{\dot{A}^2_3} \over {A^2_3}}\Bigr)\Bigr].
\end{eqnarray}

\noindent
where $\hbox{det}A=A^1_2A^2_3A^3_1$.  Making the definitions

\begin{eqnarray}
\label{MAKINGTHE5}
\Pi_f=\lambda_{ff}(\hbox{det}A);~~X^1=\hbox{ln}\Bigl({{A^3_1} \over {a_0}}\Bigr);~~X^2=\hbox{ln}\Bigl({{A^1_2} \over {a_0}}\Bigr);~~
X^3=\hbox{ln}\Bigl({{A^2_3} \over {a_0}}\Bigr),
\end{eqnarray}

\noindent
this yields the canonical commutation relations 

\begin{eqnarray}
\label{DEFINITION5}
\bigl[X^f(x,t),\Pi_f(y,t)\bigr]=\delta^f_g\delta^{(3)}(x,y).
\end{eqnarray}

\noindent
Hence we have obtained globally holonomic coordinates on the kinematic phase space $\Omega_{Kin}$ of the instanton representation, even though such coordinates do not exist on the full phase 
space $\Omega$.  Equation (\ref{MAKINGTHE5}) provides the degrees of freedom which can be used for quantization of the full theory.  This can also be written in the form

\begin{eqnarray}
\label{COINS5}
\bigl[A^f_f(x,t),\Pi_g(y,t)\bigr]=\delta^f_gA^g_g\delta^{(3)}(x,y),
\end{eqnarray}

\noindent
which are affine commutation relations.  We will refer to such configurations, where $A^a_i$ is nondegenerate and has three nonzero entries, as quantizable.  In the original Ashtekar variables this corresponds to canonical commutation relations

\begin{displaymath}
\Biggl[
\left(\begin{array}{ccc}
0 & A^1_2(x,t) & 0\\
0 & 0 & A^2_3(x,t)\\
A^3_1(x,t) & 0 & 0\\
\end{array}\right),
\left(\begin{array}{ccc}
0 & \widetilde{\sigma}^1_2(y,t) & 0\\
0 & 0 & \widetilde{\sigma}^2_3(y,t)\\
\widetilde{\sigma}^3_1(y,t) & 0 & 0\\
\end{array}\right)\Biggr]
=\left(\begin{array}{ccc}
0 & 1 & 0\\
0 &0 & 1\\
1 & 0 & 0\\
\end{array}\right)
\delta^{(3)}(x,y),
\end{displaymath}

\subsection{Sixth quantizable configuration}

The sixth case, involving the odd permutations, is given by

\begin{displaymath}
A^a_i=
\left(\begin{array}{ccc}
0 & 0 & A^1_3\\
A^2_1 & 0 & 0\\
0 & A^3_2 & 0\\
\end{array}\right);~~
B^i_a=
\left(\begin{array}{ccc}
\partial_2A^1_3 & A^3_2A^1_3 & -\partial_3A^3_2\\
-\partial_1A^1_3 & \partial_3A^2_1 & A^1_3A^3_1\\
A^2_1A^3_2 & -\partial_2A^2_1 & \partial_1A^3_2\\
\end{array}\right);
\end{displaymath}

\begin{displaymath}
B^i_e\dot{A}^a_i=
\left(\begin{array}{ccc}
A^3_2A^1_3\dot{A}^2_1 & -(\partial_3A^3_2)\dot{A}^3_2 & (\partial_2A^1_3)\dot{A}^1_3\\
(\partial_3A^2_1)\dot{A}^2_1 & A^1_3A^2_1\dot{A}^3_2 & -(\partial_1A^1_3)\dot{A}^1_3\\
-(\partial_2A^2_1)\dot{A}^2_1 & (\partial_1A^3_2)\dot{A}^3_2 & A^2_1A^3_2\dot{A}^1_3
\end{array}\right)
.
\end{displaymath}

\noindent
Upon contraction with a diagonal CDJ matrix $\Psi_{ae}=\delta_{ae}\lambda_{ee}$ this leads to the canonical structure

\begin{eqnarray}
\label{CONTRACT5}
\lambda_{11}A^3_2A^1_3\dot{A}^2_1+\lambda_{22}A^1_3A^2_1\dot{A}^3_2+\lambda_{33}A^2_1A^3_2\dot{A}^1_3\nonumber\\
=(A^3_2A^1_3A^2_1)\Bigl[\lambda_{11}\Bigl({{\dot{A}^2_1} \over {A^2_1}}\Bigr)+\lambda_{22}\Bigl({{\dot{A}^3_2} \over {A^3_2}}\Bigr)+\lambda_{33}\Bigl({{\dot{A}^1_3} \over {A^1_3}}\Bigr)\Bigr],
\end{eqnarray}

\noindent
where $\hbox{det}A=A^3_2A^1_3A^2_1$.  Making the definitions

\begin{eqnarray}
\label{MAKINGTHE6}
\Pi_f=\lambda_{ff}(\hbox{det}A);~~X^1=\hbox{ln}\Bigl({{A^2_1} \over {a_0}}\Bigr);~~X^2=\hbox{ln}\Bigl({{A^3_2} \over {a_0}}\Bigr);~~
X^3=\hbox{ln}\Bigl({{A^1_3} \over {a_0}}\Bigr),
\end{eqnarray}

\noindent
this yields the canonical commutation relations 

\begin{eqnarray}
\label{DEFINITION6}
\bigl[X^f(x,t),\Pi_f(y,t)\bigr]=\delta^f_g\delta^{(3)}(x,y).
\end{eqnarray}

\noindent
Hence we have obtained globally holonomic coordinates on the kinematic phase space $\Omega_{Kin}$ of the instanton representation, even though such coordinates do not exist on the full phase 
space $\Omega$.  Equation (\ref{MAKINGTHE5}) provides the degrees of freedom which can be used for quantization of the full theory.  This can also be written in the form

\begin{eqnarray}
\label{COINS6}
\bigl[A^f_f(x,t),\Pi_g(y,t)\bigr]=\delta^f_gA^g_g\delta^{(3)}(x,y),
\end{eqnarray}

\noindent
which are affine commutation relations.  We will refer to such configurations, where $A^a_i$ is nondegenerate and has three nonzero entries, as quantizable.  In the original Ashtekar variables this corresponds to canonical commutation relations

\begin{displaymath}
\Biggl[
\left(\begin{array}{ccc}
0 & 0 & A^1_3(x,t)\\
A^2_1(x,t) & 0 & 0\\
0 & A^3_2(x,t) & 0\\
\end{array}\right),
\left(\begin{array}{ccc}
0 & 0 & \widetilde{\sigma}^1_3(y,t)\\
\widetilde{\sigma}^2_1(y,t) & 0 & 0\\
0 & \widetilde{\sigma}^3_2(y,t) & 0\\
\end{array}\right)\Biggr]
=\left(\begin{array}{ccc}
0 & 0 & 1\\
1 &0 & 0\\
0 & 1 & 0\\
\end{array}\right)
\delta^{(3)}(x,y).
\end{displaymath}

\par
\indent
We have identified six distinct configurations of the configuration space $\Gamma_{Inst}$ for which a well-defined canonical structure can be defined.  Note, while the spatial gradients do not appear in these configurations, that they correspond to the full theory and not minisuperspace.\footnote{Hence for each of these configurations the components of the connection $A^a_i=A^a_i(x)$ can be chosen differently at each spatial point, defining three continuous functions of position.  Upon quantization, one would be quantizing the infinite dimensional spaces of field theory and not quantum mechanics, as in minisuperspace.}  Note that these are not simply repetitions of the same configuration, since they correspond to different specific combinations of Ashtekar connection components $A^a_i$ preferentially selected by the momentum space variables $\lambda_f$.  This allows $X^{ae}$ to be well-defined as a canonical variable, and will form the basis for quantization of the full theory.\footnote{As $\Psi_{ae}\equiv\delta_{af}\lambda_f$ will correspond to the diagonal matrix of eigenvalues, we will restrict our quantization to spacetimes of Petrov type $I$, $D$ and $O$.}

\subsection{Additional configurations}

\noindent
Another interesting configuration containing three D.O.F. is given by an Ashtekar connection $A^a_i$ where the diagonal elements are zero

\begin{displaymath}
A^a_i=
\left(\begin{array}{ccc}
0 & A^1_2 & A^1_3\\
A^2_1 & 0 & A^2_3\\
A^3_1 & A^3_2 & 0\\
\end{array}\right);
\end{displaymath}

\begin{displaymath}
B^i_e=
\left(\begin{array}{ccc}
\partial_2A^1_3-\partial_3A^1_2-A^2_3A^3_2 & \partial_2A^2_3+A^3_2A^1_3 & -\partial_3A^3_2+A^1_2A^2_3\\
-\partial_1A^1_3+A^2_3A^3_1 & \partial_3A^2_1-\partial_1A^2_3-A^3_1A^1_3 & \partial_3A^1_3+A^1_3A^2_1\\
\partial_1A^1_2+A^2_1A^3_2 & -\partial_2A^2_1+A^3_1A^1_2 & \partial_1A^3_2-\partial_2A^3_1-A^1_2A^2_1\\
\end{array}\right);
\end{displaymath}

\noindent
Each component of $B^i_a$ contains spatial gradients, which is problematic for the quantization.  One may circumvent this by restricting oneself to spatially homogeneous connections, which yields a magnetic field of

\begin{displaymath}
B^i_a=
\left(\begin{array}{ccc}
-A^2_3A^3_2 & A^3_2A^1_3 & A^1_2A^2_3\\
A^2_3A^3_1 & -A^3_1A^1_3 & A^1_3A^2_1\\
A^2_1A^3_2 & A^3_1A^1_2 & -A^1_2A^2_1\\
\end{array}\right).
\end{displaymath}

\noindent
The corresponding symplectic structure for the instanton representation is given by

\begin{displaymath}
B^i_e\dot{A}^a_i=
\left(\begin{array}{ccc}
A^3_2A^1_3\dot{A}^2_1+A^1_2A^2_3\dot{A}^3_1 & -A^2_3A^3_2\dot{A}^1_2+A^1_2A^2_3\dot{A}^3_2 & -A^2_3A^3_2\dot{A}^1_3+A^3_2A^1_3\dot{A}^2_3\\
-A^3_1A^1_3\dot{A}^2_1+A^1_3A^2_1\dot{A}^3_1 & A^2_3A^3_1\dot{A}^1_2+A^1_3A^2_1\dot{A}^3_2 & A^2_3A^3_1\dot{A}^1_3-A^3_1A^1_3\dot{A}^2_3\\
A^3_1A^1_2\dot{A}^2_1-A^1_2A^2_1\dot{A}^3_1 & A^2_1A^3_2\dot{A}^1_2+A^3_1A^1_2\dot{A}^2_3 & -A^2_1A^3_2\dot{A}^1_3+A^3_1A^1_2\dot{A}^2_3\\
\end{array}\right)
\end{displaymath}

\noindent
There is no obvious way to obtain a canonical structure using the off-diagonal terms.  However, for symmetric connections $A^a_i$, namely

\begin{eqnarray}
\label{SYEEM}
A^1_2=A^2_1;~~A^2_3=A^3_2;~~A^3_1=A^1_3,
\end{eqnarray}

\noindent
the diagonal terms in $B^i_a\dot{A}^a_i$ contain total time derivatives.  Upon contraction with a diagonal CDJ matrix we obtain

\begin{eqnarray}
\label{CONTRACT5}
\lambda_{11}A^2_3{d \over {dt}}(A^3_1A^1_2)+\lambda_{22}A^3_1{d \over {dt}}(A^1_2A^2_3)+\lambda_{33}A^1_2{d \over {dt}}(A^2_3A^3_1)\nonumber\\
=(A^1_2A^2_3A^3_1)\Bigl[\lambda_{11}\Bigl({{{d \over {dt}}(A^3_1A^1_2)} \over {A^3_1A^1_2}}\Bigr)
+\lambda_{22}\Bigl({{{d \over {dt}}(A^1_2A^2_3)} \over {A^1_2A^2_3}}\Bigr)
+\lambda_{33}\Bigl({{{d \over {dt}}(A^2_3A^3_1)} \over {A^2_3A^3_1}}\Bigr)\Bigr].
\end{eqnarray}

\noindent
where $\hbox{det}A=A^1_2A^2_3A^3_1$.  Hence, even though $A^a_i$ is nondiagonal, it induces diagonal canonical variables which can be used for quantization albeit in minisuperspace when one defines variables

\begin{eqnarray}
\label{WHENONE}
X^1=A^3_1A^1_2;~~X^2=A^1_2A^2_3;~~X^3=A^2_3A^3_1;~~\Pi_f=\lambda_{ff}(\hbox{det}A),
\end{eqnarray}

\noindent
where $\hbox{det}A=A^1_2A^2_3A^3_1$.  Then the following relation holds $[X^f,\Pi_g]=\delta^f_g$ for minisuperspace.

\subsection{Relation to the metric theory}

\noindent
The phase space in metric variables is given by $(h_{ij},\pi^{ij})$, which at the unconstrained level consists of twelve phase space degrees of freedom.  Upon implementation of the diffeomorphism constraint $H_i$, we should have a cotangent bundle structure with six phase space degrees of freedom.  This constitutes the analogue of the kinematic phase space $\Omega_{Kin}$ of the instanton representation, where the dynamics of the Hamiltonian constraint can be implemented to yield the physical phase space $\Omega_{Phys}$.  The Dirac method to quantize the metric representation would be to write the canonical commutation relations on the full unconstrained phase space

\begin{eqnarray}
\label{UNCONSTRAINED}
\bigl[h_{ij}(x,t),\pi^{mn}(y,t)\bigr]=\delta^m_i\delta^n_j\delta^{(3)}(x,y).
\end{eqnarray}

\noindent
On the reduced phase space corresponding to invariance under spatial diffeomorphisms, only three degrees of freedom from (\ref{UNCONSTRAINED}) should be quantized.  On this space $\Omega_{Kin}$ the 
metric $h_{ij}$ should be diagonal,\footnote{This is the only way to obtain a Riemannian metric, of signature $(1,1,1)$ on three dimensional space, which incidentally corresponds to a spacetime metric of signature $(-1,1,1,1)$.} which in turn implies that the conjugate momentum $\pi^{ij}$ must also be diagonal in order to preserve a cotangent bundle structure.  Equation (\ref{UNCONSTRAINED}) is invariant under $SO(3)$ transformations, an observation which we can exploit in obtaining $\Omega_{Kin}$  For nondegenerate variables we can write

\begin{eqnarray}
\label{UNCONSTRAINED1}
\bigl[O_{ik}h_kO^T_{kj},U^{ml}\pi^l(U^T)^{ln}\bigr],
\end{eqnarray}

\noindent
where $O$ and $U$ are orthogonal matrices.  It is shown in Paper IV that the $SO(3,C)$ angles used to diagonalize the CDJ matrix $\Psi_{ae}$ can at the canonical level be considered ignorable.  The analogue for (\ref{UNCONSTRAINED1}) would be to multiply by $O^T_{i^{\prime}i}(U^T)^{m^{\prime}m}O_{jj^{\prime}}U^{nn^{\prime}}$.  If $O=U$, then the metric and its conjugate momentum are diagonalized by the same $SO(3)$ transformation, which transforms the relations into

\begin{eqnarray}
\label{UNCONSTRAINED2}
\delta_{i^{\prime}k}\delta_{j^{\prime}k}\delta^{n^{\prime}l}\delta^{m^{\prime}l}\bigl[h_k(x,t),\pi^l(y,t)\bigr]=\delta^{m^{\prime}}_{i^{\prime}}\delta^{n^{\prime}}_{j^{\prime}}\delta^{(3)}(x,y).
\end{eqnarray}

\noindent
Since only the diagonal configurations contribute, then the Kronecker deltas can be dropped and the canonical commutation relations on the kinematic phase space can be written as

\begin{eqnarray}
\label{UNCONSTRAINED3}
\bigl[h_k(x,t),\pi^l(y,t)\bigr]=\delta^l_k\delta^{(3)}(x,y).
\end{eqnarray}

\noindent
The diagonal metric can be mapped directly to the quantizable instanton configurations, which yields the six possibilities on the kinematic phase space

\begin{displaymath}
h^{(1)}_{ij}=
\left(\begin{array}{ccc}
\widetilde{\sigma}^1_1\widetilde{\sigma}^1_1 & 0 & 0\\
0 & \widetilde{\sigma}^2_2\widetilde{\sigma}^2_2 & 0\\
0 & 0 & \widetilde{\sigma}^3_3\widetilde{\sigma}^3_3\\
\end{array}\right);~~
h^{(2)}_{ij}=
\left(\begin{array}{ccc}
\widetilde{\sigma}^1_1\widetilde{\sigma}^1_1 & 0 & 0\\
0 & \widetilde{\sigma}^2_3\widetilde{\sigma}^2_3 & 0\\
0 & 0 & \widetilde{\sigma}^3_2\widetilde{\sigma}^3_2\\
\end{array}\right);
\end{displaymath}

\begin{displaymath}
h^{(3)}_{ij}=
\left(\begin{array}{ccc}
\widetilde{\sigma}^1_3\widetilde{\sigma}^1_3 & 0 & 0\\
0 & \widetilde{\sigma}^2_2\widetilde{\sigma}^2_2 & 0\\
0 & 0 & \widetilde{\sigma}^3_1\widetilde{\sigma}^3_1\\
\end{array}\right);
h^{(4)}_{ij}=
\left(\begin{array}{ccc}
\widetilde{\sigma}^1_2\widetilde{\sigma}^1_2 & 0 & 0\\
0 & \widetilde{\sigma}^2_1\widetilde{\sigma}^2_1 & 0\\
0 & 0 & \widetilde{\sigma}^3_3\widetilde{\sigma}^3_3\\
\end{array}\right);
\end{displaymath}

\begin{displaymath}
h^{(5)}_{ij}=
\left(\begin{array}{ccc}
\widetilde{\sigma}^1_2\widetilde{\sigma}^1_2 & 0 & 0\\
0 & \widetilde{\sigma}^2_3\widetilde{\sigma}^2_3 & 0\\
0 & 0 & \widetilde{\sigma}^3_1\widetilde{\sigma}^3_1\\
\end{array}\right);
h^{(6)}_{ij}=
\left(\begin{array}{ccc}
\widetilde{\sigma}^2_1\widetilde{\sigma}^2_1 & 0 & 0\\
0 & \widetilde{\sigma}^3_2\widetilde{\sigma}^3_2 & 0\\
0 & 0 & \widetilde{\sigma}^1_3\widetilde{\sigma}^1_3\\
\end{array}\right)
.
\end{displaymath}

\noindent
The question then arises as to whether a diagonal 3-metric on the kinematic phase space excludes off-diagonal configurations for the spacetime metric $g_{\mu\nu}$.  The answer is no, since any off-diagonal parts of $g_{\mu\nu}$ can be attributed to the existence of a nonvanishing shift vector $N^i$.  The purpose of the diffeomorphism constraint is to eliminate this contribtution in bringing us to the kinematic phase space.  Such off-diagonal terms, as we have shown, should not be on the same footing as the diagonal terms corresponding to the intrinsic $SO(3,C)$ frame.

\newpage

\section{Self-dual Weyl curvature tensor}

One of the outstanding issues in quantum gravity is to get a handle on the manifestation of the quantum theory in the classical limit in terms of measurable quantities.  We have shown that there exist configurations on the kinematic phase space of the instanton representation which it is possible to quantize.  The purpose of the next two sections will be to provide the interpretation of what will be quantized, in terms of directly measurable quantities of physical significance for general relativity in the semiclassical limit.  This requires some introductory material on the Weyl curvature tensor and its relation to the physical degrees of freedom.\par
\indent
The instanton representation of Plebanski gravity implies that the physical degrees of freedom of gravity may be encoded within the self dual part of the Weyl curvature tensor, denoted $Weyl$.  The Weyl curvature 
tensor $C_{\mu\nu\rho\sigma}$ is the traceless part of the Riemann curvature $R_{\mu\nu\rho\sigma}$, given by \cite{WHEELER}

\begin{eqnarray}
\label{REIMAN}
R_{\mu\nu\rho\sigma}={1 \over 6}\bigl(g_{\rho\nu}g_{\mu\sigma}-g_{\rho\mu}g_{\nu\sigma}\bigr)R+C_{\mu\nu\rho\sigma}\nonumber\\
+{1 \over 2}\bigl(g_{\rho\mu}R_{\nu\sigma}-g_{\rho\nu}R_{\mu\sigma}
-g_{\sigma\mu}R_{\nu\rho}+g_{\sigma\nu}R_{\mu\rho}\bigr).
\end{eqnarray}

\noindent
The Weyl curvature tensor describes the nonlocal effects of radiation on curvature not including matter fields and encodes the algebraic classification of spacetime.  
Equation (\ref{REIMAN}) can be decomposed into electric and magnetic parts $E_{\mu\nu}$ and $B_{\mu\nu}$ with respect to an observer with 4-velocity $u^{\mu}$ tangent to a congruence of timelike integral curves as \cite{MACCALLUM}

\begin{eqnarray}
\label{REIMAN1}
Q_{\mu\rho}=(C_{\mu\nu\rho\sigma}+i^{*}C_{\mu\nu\rho\sigma})u^{\nu}u^{\sigma}.
\end{eqnarray}

\noindent
Note that $Q_{\mu\nu}u^{\nu}=0$, namely that $Q_{\mu\nu}$ lives in the three dimensional space orthogonal to $u^{\mu}$.  For $u^{\mu}=\delta^{\mu}_0$ the tensor $Q_{\mu\nu}$ is purely spatial, and can be written as a symmetric traceless three by three matrix $Q_{ij}=E_{ij}+iB_{ij}$.  This is given for vanishing cosmological constant by \cite{THREEONE}

\begin{eqnarray}
\label{MAGNETICPART}
E_{ij}=R_{ij}-\pi_i^k\pi_{kj}+(\hbox{tr}\pi)\pi_{ij};~~B_{ij}=\epsilon_i^{kl}\nabla_k\pi_{lj},
\end{eqnarray}

\noindent
where $(h_{ij},\pi^{ij})$ are the 3-metric of $\Sigma$ and its conjugate momentum, where $\Sigma$ represents 3-space and $\nabla_i$ is the three dimensional Levi--Civita connection for $h_{ij}$, with 
Ricci curvature $R_{ij}v^j=-2\nabla_{[i}\nabla_{j]}v^j$ and Ricci scalar $R=R_{ij}h^{ij}$.  When $Q_{ij}$ is diagonalizable, it can be written in a canonical frame such that it is diagonal

\begin{displaymath}
Q_{ij}=
\left(\begin{array}{ccc}
\lambda_1 & 0 & 0\\
0 & \lambda_2 & 0\\
0 & 0 & \lambda_3\\
\end{array}\right)
.
\end{displaymath}

\noindent
The eigenvalues of $Q_{ij}$ define the algebraic properties of spacetime which are invariant under coordinate transformations and the choice of a tetrad frame.  At most two eigenvalues are independent due to the tracefree condition 

\begin{eqnarray}
\label{TRACEFREE}
Q^i_i=\lambda_1+\lambda_2+\lambda_3=0.
\end{eqnarray}

\par
\indent
The Petrov classification distinguishes between algebraically general (Petrov Type I) and algebraically special spacetimes (Petrov types II, III, N, D, O) according to the degeneracy of eigenvalues and eigenvectors of $Q_{ij}$.  One defines invariants $I$ and $J$, given by \cite{KASNER}

\begin{eqnarray}
\label{INVARIANTS}
I={1 \over 2}\hbox{tr}Q^2={1 \over 2}\bigl(\lambda_1^2+\lambda_2^2+\lambda_3^2\bigr);~~J={1 \over 6}\hbox{tr}Q^3={1 \over 6}\bigl(\lambda_1^3+\lambda_2^3+\lambda_3^3\bigr),
\end{eqnarray}

\noindent
and finds the eigenvalues from the characteristic equation for $Q_{ij}$

\begin{eqnarray}
\label{EIEMM}
\lambda^3-I\lambda+2J=0.
\end{eqnarray}

\noindent
The real and the imaginary parts of $I$ are given by \cite{ELECTRIC}

\begin{eqnarray}
\label{INVARIAIRII}
Re[I]={1 \over 2}\bigl(E_{ij}E^{ij}-B_{ij}B^{ij}\bigr);~~Im[I]={1 \over 2}E_{ij}B^{ij},
\end{eqnarray}

\noindent
which is reminiscent of the radiative invariants $\vec{E}\cdot\vec{E}-\vec{B}\cdot\vec{B}$ and $\vec{E}\cdot\vec{B}$ for electromagnetism.  From the invariants $I$ and $J$ can be defined a specialty index $S$, given by

\begin{eqnarray}
\label{SPECIAL}
S={{27J^2} \over {I^3}},
\end{eqnarray}

\noindent
where $S=1$ for algebraically special spacetimes.  For Petrov types III, N and O the invariants $I=J=0$, and for Petrov types II and D they are nontrivial.

\subsection{Two component spinor SL(2,C) formalism}

To place the CDJ matrix $\Psi_{ae}$ into context, it is instructive to establish its relation to the $SL(2,C)$ formalism of GR \cite{GROUP}.  Define two component spinors $\eta^A$ and their complex conjugates $\overline{\eta}^A$, where $A$ and $A^{\prime}$ respectively denote left handed and right handed $SL(2,C)$ spinorial indices.\footnote{Spinorial indices take on the values $0$ and $1$ for both primed and unprimed indices.  Note that $SL(2,C)$ is the covering group for $SO(3,C)$, and we regard the CDJ matrix $\Psi_{ae}$ as taking values in two copies of the left-handed $SO(3,C)$.}  These indices are raised and lowered by the two dimensional Levi--Civita symbol $\epsilon_{AB}$, where

\begin{eqnarray}
\label{LEVI}
\eta^A=\epsilon^{AB}\eta_B;~~\eta_B=\eta^A\epsilon_{AB};~~\overline{\eta}^{A^{\prime}}=\epsilon^{A^{\prime}B^{\prime}}\eta_{B^{\prime}};~~\eta_{B^{\prime}}=\eta^{A^{\prime}}\epsilon_{A^{\prime}B^{\prime}}.
\end{eqnarray}

\noindent
To connect the internal spinor space to objects containing world indices $\mu$, one can define $\forall{x}\in{M}$ a set of soldering forms $\sigma^{\mu}_{AA^{\prime}}$, which in a Cartesian coordinate system may take on the matrix form

\begin{displaymath}
\sigma^0_{AA^{\prime}}={i \over {\sqrt{2}}}
\left(\begin{array}{cc}
1 & 0\\
0 & 1\\
\end{array}\right);~~
\sigma^1_{AA^{\prime}}={i \over {\sqrt{2}}}
\left(\begin{array}{cc}
0 & 1\\
1 & 0\\
\end{array}\right);
\end{displaymath}

\begin{displaymath}
\sigma^2_{AA^{\prime}}={i \over {\sqrt{2}}}
\left(\begin{array}{cc}
0 & -i\\
i & 0\\
\end{array}\right);~~
\sigma^3_{AA^{\prime}}={i \over {\sqrt{2}}}
\left(\begin{array}{cc}
1 & 0\\
0 & -1\\
\end{array}\right).
\end{displaymath}

\noindent
The soldering forms $\sigma^{\mu}_{AA^{\prime}}$ define an isomorphism between 4-vectors and spinorial pairs in the $({1 \over 2},{1 \over 2})$ representation at each point of $M$, and a 4-vector $V_{\mu}$ decomposes as

\begin{eqnarray}
\label{FOURVEC}
V_{AA^{\prime}}=V_{\mu}\sigma^{\mu}_{AA^{\prime}};~~V^{\mu}=\sigma^{\mu}_{AA^{\prime}}V^{AA^{\prime}}.
\end{eqnarray}

\indent
Tensors of multiple rank decompose by generalization of (\ref{FOURVEC}) as

\begin{eqnarray}
\label{FOURTENS}
V_{A_1A_1^{\prime}A_2A_2^{\prime}\dots{A}_nA_n^{\prime}}=V_{\mu_1\mu_2\dots\mu_n}\sigma^{\mu_1}_{A_1A_1^{\prime}}\sigma^{\mu_2}_{A_2A_2^{\prime}}\dots\sigma^{\mu_n}_{A_nA_n^{\prime}}
\end{eqnarray}
 
\noindent
and similarly

\begin{eqnarray}
\label{FOURTENS1}
V_{\mu_1\mu_2\dots\mu_n}=\sigma_{\mu_1}^{A_1A_1^{\prime}}\sigma_{\mu_2}^{A_2A_2^{\prime}}\dots\sigma_{\mu_n}^{A_nA_n^{\prime}}V_{A_1A_1^{\prime}A_2A_2^{\prime}\dots{A}_nA_n^{\prime}}.
\end{eqnarray}

\par
\indent
Any pair of left handed null two component spinors $n_A$ and $l_A$ in $M$ satisfying the normalization conditions 

\begin{eqnarray}
\label{NULL}
n^An_A=l^Al_A=0;~~n^Al_A=1,
\end{eqnarray}

\noindent
in conjunction with the soldering form $\sigma^{\mu}_{AA^{\prime}}$ defines a null tetrad

\begin{eqnarray}
\label{FOURVEC4}
l^{\mu}=\sigma^{\mu}_{AA^{\prime}}l^A\overline{l}^{A^{\prime}};~~
n^{\mu}=\sigma^{\mu}_{AA^{\prime}}n^A\overline{n}^{A^{\prime}};~~
m^{\mu}=\sigma^{\mu}_{AA^{\prime}}n^A\overline{l}^{A^{\prime}};~~
\overline{m}^{\mu}=\sigma^{\mu}_{AA^{\prime}}l^A\overline{n}^{A^{\prime}},
\end{eqnarray}

\noindent
such that

\begin{eqnarray}
\label{FOURVEC41}
l^{\mu}l_{\mu}=m_{\mu}m^{\mu}=\overline{m}_{\mu}\overline{m}^{\mu}=n^{\mu}n_{\mu}=0;\nonumber\\
l_{\mu}m^{\mu}=l_{\mu}\overline{m}^{\mu}=n_{\mu}m^{\mu}=n_{\mu}\overline{m}^{\mu}=0;\nonumber\\
l_{\mu}n^{\mu}=-m_{\mu}\overline{m}^{\mu}=1.
\end{eqnarray}

\noindent
The null vectors $l^{\mu}$ and $n^{\nu}$ are real and span a time-like 2-plane in $T_p(M)$, the tangent space at each point of spacetime $M$.  The null vectors $m^{\mu}$ and $\overline{m}^{\mu}$ are complex, and span the orthogonal space-like 2-plane in $T_p(M)$.   The tetrad $(l^{\mu},n^{\mu}m^{\mu},\overline{m}^{\mu})$ is useful in the Penrose approach to GR \cite{SWALD}, which is suited to characterizing the radiation properties of spacetime \cite{NULL}.\par
\indent  
The spinors $(n_A,l_A)$ induce a basis $\eta^a_{AB}$ in spin space \cite{GROUP}, one such basis given by 

\begin{eqnarray}
\label{FORUVEC5}
\eta^1_{AB}=\sqrt{2}il_{(A}n_{B)};~~\eta^2_{AB}={i \over {\sqrt{2}}}\bigl(l_Al_B+n_An_B\bigr);~~\eta^3_{AB}={i \over {\sqrt{2}}}\bigl(l_Al_B-n_An_B\bigr),
\end{eqnarray}

\noindent
where 

\begin{eqnarray}
\label{WHERE}
\eta^a_{AB}\eta_f^{AB}=\delta^a_f.
\end{eqnarray}

\noindent
Thsese objects define an isomorphism between internal indices $a=(1,2,3)$ and symmetric $SU(2)$ index pairs $(00)$, $(01)$ and $(10)$.  Any dyad can be expressed in the basis (\ref{FORUVEC5}) as 

\begin{eqnarray}
\label{EXPRESS}
\phi_{AB}=\sum_{m=1}^3\chi_m(\eta^m)_{AB},
\end{eqnarray}

\noindent
where $\chi_m$ are the components.  A $SL(2,C)$ transformation $g$, acting on the column vector $(l_A,n_A)$, given by

\begin{displaymath}
g=
\left(\begin{array}{cc}
a & b\\
c & d\\
\end{array}\right)
;~~ad-bc=1,
\end{displaymath}

\noindent
induces a transformation of the basis and the corresponding components

\begin{displaymath}
\left(\begin{array}{c}
\phi^{\prime}_1\\
\phi^{\prime}_2\\
\phi^{\prime}_3\\
\end{array}\right)
=
\left(\begin{array}{ccc}
a^2 & 2ab & b^2\\
ac & bc+ad & bd\\
c^2 & 2cd & d^2\\
\end{array}\right)
\left(\begin{array}{c}
\phi_1\\
\phi_2\\
\phi_3\\
\end{array}\right)
.
\end{displaymath}

\noindent
The basis (\ref{FORUVEC5}) also induces an orthonormal basis of completely symmetric four-spinors, given by \cite{GROUP}

\begin{eqnarray}
\label{SYMMETRIC}
\eta^0_{ABCD}={1 \over {\sqrt{2}}}\bigl(l_Al_Bl_Cl_D+n_An_Bn_Cn_D\bigr);\nonumber\\
\eta^1_{ABCD}=\sqrt{2}i\bigl(l_{(A}l_Bl_Cn_{D)}+l_{(A}n_Bn_Cn_{D)}\bigr);\nonumber\\
\eta^2_{ABCD}=\sqrt{6}l_{(A}l_Bn_Cn_{D)};\nonumber\\
\eta^3_{ABCD}=\sqrt{2}\bigl(l_{(A}l_Bl_Cn_{D)}-l_{(A}n_Bn_Cn_{D)};\nonumber\\
\eta^4_{ABCD}={i \over {\sqrt{2}}}\bigl(l_Al_Bl_Cl_D-n_An_Bn_Cn_D\bigr),
\end{eqnarray}

\noindent
satisfying orthonormality relations

\begin{eqnarray}
\label{SYMMETRIC1}
\eta^{\alpha}_{ABCD}(\eta^{\beta})^{ABCD}=\delta^{\alpha\beta}.
\end{eqnarray}

\subsection{Principal null directions of spacetime}

By application of (\ref{FOURTENS}) and (\ref{FOURTENS1}), one may decompose the Riemann curvature tensor into $SL(2,C)$ indices as

\begin{eqnarray}
\label{FOURVEC1}
R_{\mu\nu\rho\sigma}\sigma^{\mu}_{AA^{\prime}}\sigma^{\nu}_{BB^{\prime}}\sigma^{\rho}_{CC^{\prime}}\sigma^{\sigma}_{DD^{\prime}}
=\epsilon_{A^{\prime}B^{\prime}}\epsilon_{C^{\prime}D^{\prime}}\psi_{ABCD}+\dots,
\end{eqnarray}

\noindent
where $\psi_{ABCD}=\psi_{(ABCD)}$ is $Weyl$, the self-dual part of the Weyl curvature tensor, and the dots signify the remaining components which will not concern us in this paper.  
Using the basis (\ref{SYMMETRIC}), $Weyl$ can be written as

\begin{eqnarray}
\label{SYMMETRIC2}
\psi_{ABCD}=2\sum_{f=0}^4\Psi_{\alpha}\eta^\alpha_{ABCD},
\end{eqnarray}

\noindent
where $\Psi_{\alpha}$ are defined as the Weyl scalars.  In a suitable adapted frame, the Weyl scalars may be shown to admit the following physical interpretations in vacuum spacetimes: (i) $\Psi_0$ and $\Psi_4$ are transverse components of gravitational radiation propagating in the $l^{\mu}$ and the $n^{\mu}$ directions respectively. (ii) $\Psi_1$ and $\Psi_3$ are longitudinal components propagating in the $l^{\mu}$ and the $n^{\mu}$ directions 
respectively. (iii) $\Psi_2$ is a Coulombic component.\par
\indent
The principal null directions of spacetime can be computed directly by performing an $SL(2,C)$ transformation to eliminate $\Psi_4(\Psi_0)$, which leaves $l^{\mu}(n^{\mu})$ as the principal null direction.  For example a null rotation 
which keeps $l^A$ invariant transforms $\Psi_4$ into 

\begin{eqnarray}
\label{COMPUTEIT2}
\Psi^{\prime}_4=\Psi_4+4\Psi_3z+6\Psi_2z^2+4\Psi_1z^3+\Psi_0z^4.
\end{eqnarray}

\noindent
The condition for $l^{\mu}$ to be a principal null direction is that $\Psi^{\prime}_4=0$.  This yields a quartic polynomial equation in $z$, which is also given by

\begin{eqnarray}
\label{WEYLBASIS}
\Psi_{ABCD}\xi^A\xi^B\xi^C\xi^D=0.
\end{eqnarray}

\noindent
Equation (\ref{COMPUTEIT2}) in general has four roots, and the multiplicity of each principal null direction is the same as the multiplicity of the 
corresponding root.  The roots $z_i$ for $i=1,\dots{4}$ can be parametrized as \cite{THREEONE}

\begin{eqnarray}
\label{COMPITEITT}
z_i=\hbox{tan}(\theta_i/2)e^{-i\phi_i},
\end{eqnarray}

\noindent
which can be put in one-to-one correspondence with points on the two-sphere by stereographic projection of $z_i$.  The principal null directions are given by

\begin{eqnarray}
\label{COMPETITEE}
P^a_{(i)}=\hbox{cos}\theta_i\hat{z}^a+\hbox{sin}\theta_i\hbox{cos}\phi_i\hat{x}^a+\hbox{sin}\theta_i\hbox{sin}\phi_i\hat{y}^a,
\end{eqnarray}

\noindent
where $(\theta_i,\phi_i)$ coordinatize angular position on the two-sphere.  The number and multiplicity of principal null directions determines the Petrov classification of $Weyl$.  The Petrov classification scheme then is as follows

\begin{eqnarray}
\label{SCHEME}
Type~I~(1,1,1,1);~~Type~II~(2,1,1);~~Type~D~(2,2);\nonumber\\
Type~III~(3,1);~~Type~N~(4);~~Type~O (Conformally~flat).
\end{eqnarray}

\noindent
In brackets we have indicated the multiplicity of PNDs within each category.  From $\psi_{ABCD}$ one can form the invariants 
 
\begin{eqnarray}
\label{FOURVEC2}
I={1 \over 2}\psi_{ABCD}\psi^{ABCD};~~J={1 \over 6}\psi_{ABCD}\psi^{CDEF}\psi_{EF}^{AB},
\end{eqnarray}

\noindent
which in direct analogy to (\ref{SPECIAL}) define a specialty index $S$ given by

\begin{eqnarray}
\label{FOURVEC3}
S^{\prime}={{I^3} \over {J^2}}-27.
\end{eqnarray}

\noindent

\subsection{Relation to the CDJ matrix}

\noindent
We will now establish a direct correspondence from the principal null directions of spacetime to the quantizable degrees of freedom in the instanton representation.  Using the basis (\ref{FORUVEC5}), $Weyl$ can also be decomposed into the following form

\begin{eqnarray}
\label{CANBE}
\psi_{ABCD}=\sum_{a,e=1}^3\psi_{ae}\eta^a_{AB}\eta^e_{CD},
\end{eqnarray}

\noindent
which defines a symmetric and traceless matrix $\psi_{ae}$.  The relation between $\psi_{ae}$ and the Weyl scalars in this basis $\Psi_{\alpha}$ is given by \cite{GROUP}

\begin{displaymath}
\psi_{ae}=
\left(\begin{array}{ccc}
-2\Psi_2 & i(\Psi_1+\Psi_3) & (\Psi_3-\Psi_1)\\
i(\Psi_1+\Psi_3) & {1 \over 2}(2\Psi_2+\Psi_0+\Psi_4) & {i \over 2}(\Psi_0-\Psi_4)\\
(\Psi_3-\Psi_1) & {i \over 2}(\Psi_0-\Psi_4) & {1 \over 2}(2\Psi_2-\Psi_0-\Psi_4)\\
\end{array}\right)
.
\end{displaymath}

\noindent
The invariants of $\psi_{ae}$ are given by

\begin{eqnarray}
\label{INVARIANTS}
2I=\hbox{tr}\psi^2=2\Psi_0\Psi_4-8\Psi_1\Psi_3+6\Psi_2^2;\nonumber\\
6J=\hbox{det}\Psi=2\bigl(\Psi_0\Psi_2\Psi_4-\Psi_0\Psi_3^2-\Psi_4\Psi_1^2+2\Psi_1\Psi_2\Psi_3-\Psi_2^3\bigr)
\end{eqnarray}

\noindent
which imply the characteristic equation

\begin{eqnarray}
\label{CEEDEE1}
r^3-Ir+2J=0.
\end{eqnarray}

\noindent
A Weyl spinor $\psi_{ABCD}$ of Petrov Type I, has four distinct PNDs at any point, which in a certain frame form the vertices of a disphenoid \cite{PENROSERIND}.  This disphenoid represents the intersection of a spacelike plane 
with $S^{+}$, the cone of null directions at that point.  It has been elaborated in \cite{NULL1} the relation of this dispenoid to the geometry of the roots of (\ref{CEEDEE1}).  One takes one PND from (\ref{COMPITEITT}) as the north pole of $S^{+}$ and projects stereographically the other three PND onto the extended Argand plane.  The shape of the PND on $S^{+}$ mirrors the pattern of eigenvalues of $\psi_{ae}$, fixed by (\ref{CEEDEE1}), which are the vertices of a triangle in the complex plane.\par
\indent
The roots $r_1$, $r_2$ and $r_3$ of (\ref{CEEDEE1}) depend explicitly on $I$ and $J$ from (\ref{INVARIANTS}), and the CDJ matrix $\Psi_{ae}$ is defined by the addition of a spin 0 part to $\psi_{ae}$, hence the relation

\begin{eqnarray}
\label{CEEDEE2}
\Psi^{-1}_{ae}-\delta_{ae}\varphi=\psi_{ae}, 
\end{eqnarray}

\noindent
where $\varphi={1 \over 3}(\hbox{tr}\Psi^{-1})$.  Equation (\ref{CEEDEE2}) can be inverted to yield $\Psi_{ae}=\bigl(\delta_{ae}\varphi+\psi_{ae}\bigr)^{-1}$.  Therefore, since $\psi_{ae}=\psi_{ae}(I,J)$ encodes the algebraic classification of the spacetime, it follows that the CDJ matrix $\Psi_{ae}=\Psi_{ae}(I,J)$ also encodes this algebraic classification.\par
\indent
The decomposition of $Weyl$ into electric and magnetic parts (\ref{MAGNETICPART}) in spacetime $M$ is generally known.  But also in four spacetime dimensions, there is a three dimensional vector space $\textbf{W}^{-}$ spanned by the triple of self-dual $SO(3,C)$ two forms $\Sigma^a$ \cite{BENGT}.  Expansion of (\ref{REIMAN}) with respect to $\textbf{W}^{-}$ yields a symmetric and traceless three by three matrix $\psi_{ae}$, related to the self-dual part of $Weyl$ by

\begin{eqnarray}
\label{REIIM}
C_{\mu\nu\rho\sigma}=\psi_{ae}\Sigma^a_{\mu\nu}\Sigma^e_{\rho\sigma},
\end{eqnarray}

\noindent
which can also be seen from (\ref{CANBE}).  Using (\ref{REIMAN1}), the following relation can be written 

\begin{eqnarray}
\label{REIIM1}
Q_{\mu\nu}=\psi_{ae}\Sigma^a_{\mu\rho}\Sigma^e_{\nu\sigma}u^{\rho}u^{\sigma},
\end{eqnarray}

\noindent
which relates the spacetime object $Q_{\mu\nu}$ to the purely internal object $\psi_{ae}$.  For $u^{\mu}=\delta^{\mu}_0$, the decomposition reads

\begin{eqnarray}
\label{READS}
Q_{ij}=\psi_{ae}\Sigma^a_{0i}\Sigma^e_{0j},
\end{eqnarray}

\noindent
which involves the temporal components $\Sigma^a_{0i}$ of the two forms.  Hence it is clear that $Q_{\mu\nu}$ contains the same number of D.O.F. as does $\psi_{ae}$, 
since $\Sigma^a_{\mu\nu}u^{\mu}u^{\nu}=0$.  We will regard $\psi_{ae}$ as being the fundamental object, with $Q_{ij}$ being derived upon specification specification of a self-dual two form and a choice of Lorentz 
observer.

\subsection{Quantizable degrees of freedom}

For spacetimes of Petrov 
type $I$, $D$ and $O$, $\psi_{ae}$ contains three linearly independent eigenvectors and can be diagonalized according to \cite{WEYL}.  For these spacetimes, one can perform a $SO(3,C)$ transformation 
of (\ref{CEEDEE2}), putting the symmetric part of the CDJ matrix $\Psi_{ae}$ into diagonal form   

\begin{displaymath}
\Psi_{(ae)}=(e^{\theta\cdot{T}})_{ab}
\left(\begin{array}{ccc}
\lambda & 0 & 0\\
0 & \lambda+\alpha & 0\\
0 & 0 & \lambda+\beta\\
\end{array}\right)_{bc}
(e^{-\theta\cdot{T}})_{ce}
\end{displaymath}
 
\noindent
for some $\lambda$, $\alpha$ and $\beta$.  This is a polar decomposition of $\Psi_{ae}$ using a complex orthogonal transformation parametrized by three complex angles $\vec{\theta}=(\theta^1,\theta^2,\theta^3)$.  The Hamiltonian constraint (\ref{SOLU1}) is a relation amongst the eigenvalues which is independent of the $SO(3,C)$ frame, and can be written as

\begin{eqnarray}
\label{CEEDEE4}
\Lambda+\hbox{tr}\Psi^{-1}=
\Lambda+{1 \over \lambda}+{1 \over {\lambda+\alpha}}+{1 \over {\lambda+\beta}}=0.
\end{eqnarray}

\noindent
Equation (\ref{CEEDEE4}) for $\Lambda\neq{0}$ implies the cubic equation

\begin{eqnarray}
\label{IMPLIESTHE}
\lambda(\lambda+\alpha)(\lambda+\beta)+{3 \over \Lambda}\bigl(\lambda^2+{2 \over 3}(\alpha+\beta)\lambda+{1 \over 3}\alpha\beta\bigr)=0\longrightarrow\lambda=\lambda_{\alpha,\beta}
\end{eqnarray}

\noindent
where $\lambda_{\alpha,\beta}$ are the roots.  One may use the same $SO(3,C)$ transformation to diagonalize 
both sides of (\ref{CEEDEE2}), obtaining the relations

\begin{eqnarray}
\label{CEEDEE5}
{1 \over {\lambda_{\alpha,\beta}+\alpha}}+{\Lambda \over 3}=r_1(I,J);~~
{1 \over {\lambda_{\alpha,\beta}+\beta}}+{\Lambda \over 3}=r_2(I,J);~~{1 \over {\lambda_{\alpha,\beta}}}+{\Lambda \over 3}=r_3(I,J).
\end{eqnarray}

\noindent
Equation (\ref{CEEDEE5}) is a system of three equations in two unknowns which should provide $\alpha=\alpha(I,J)$, $\beta=\beta(I,J)$ and $\lambda=\lambda(I,J)$ explicitly through the roots of (\ref{CEEDEE1}).  These roots must satisfy the relation $r_1+r_2+r_3=0$, which is equivalent to the tracelessness of $\psi_{ae}$.  By inverting (\ref{CEEDEE5}) we can express $\alpha$, $\beta$ and $\lambda$, which are directly related to the eigenvalues of $\Psi_{ae}$ directly in terms of $r_1$, $r_2$ and $r_3$

\begin{eqnarray}
\label{CEEDEE7}
\lambda_{\alpha,\beta}=\Bigl({1 \over {r_3-{\Lambda \over 3}}}\Bigr)=-\Bigl({1 \over {r_1+r_2+{\Lambda \over 3}}}\Bigr);\nonumber\\
\alpha={1 \over {r_1-{\Lambda \over 3}}}+{1 \over {r_1+r_2+{\Lambda \over 3}}};~~
\beta={1 \over {r_2-{\Lambda \over 3}}}+{1 \over {r_1+r_2+{\Lambda \over 3}}}.
\end{eqnarray}

\noindent
In a solution to the initial value constraints the $SO(3,C)$ angles $\vec{\theta}$ are not arbitrary, but must satisfy the Gauss' law constraint

\begin{eqnarray}
\label{GAUSSLAW}
\textbf{w}_e\{(e^{\theta\cdot{T}})_{af}\lambda_f(e^{-\theta\cdot{T}})_{fe}\}=0
\end{eqnarray}

\noindent
for $\vec{\theta}=\vec{\theta}[\vec{\lambda};A]$ where $\lambda_f=(\lambda_{\alpha,\beta},\lambda_{\alpha,\beta}+\alpha,\lambda_{\alpha,\beta}+\beta)$ satisfying (\ref{CEEDEE4}).  Each configuration $A^a_i$ defines an equivalence class of $SO(3,C)$ frames corresponding to the eigenvalues thus chosen.  Since $\lambda_f$ are coordinate independent, then it follows that the coordinate-dependent information in the principal null directions must be 
encoded in $A^a_i$.\par
\indent
One can formulate a quantum theory of the algebraic classification of spacetime by quantizing the eigenvalues of the CDJ matrix $\Psi_{ae}$, regarded as the physical degrees of freedom of the momentum space.  We have proven the existence of configuration space variables which are canonically conjugate to the eigenvalues of $\Psi_{ae}$.\footnote{It is shown in Paper IV that the angles $\vec{\theta}$, at the canonical level, are not independent degrees of freedom and therefore should not be quantized.}  
Since the aim of quantization is to construct quantum states corresponding to the CDJ matrix, then we need a prescription for relating its invariants to the invariants of spacetime.  Starting from the CDJ matrix

\begin{eqnarray}
\label{SPACETIME}
\Psi^{-1}_{ae}=\delta_{ae}\varphi+\psi_{ae},
\end{eqnarray}

\noindent
where $\varphi=-{\Lambda \over 3}$, we have the following library of terms

\begin{eqnarray}
\label{SPACETIME1}
\hbox{tr}\Psi^{-1}=3\varphi;\nonumber\\
\hbox{tr}(\Psi^{-1}\Psi^{-1})=3\varphi^2+\hbox{tr}\psi^2=3\varphi^2+2I\equiv{M};\nonumber\\
\hbox{det}\Psi^{-1}=(\hbox{det}\Psi)^{-1}=\varphi^3-I\varphi+2J\equiv{Q}.
\end{eqnarray}

\noindent
For type $N$, $II$ and $III$ spacetimes $\psi_{ae}$ is not diagonalizable, and so we defer the quantization of such spacetimes for future study.\par
\indent
We now provide a prescription for computing the principal null directions directly from the state labels, as follows.  For the given algebraic type that one is in one must first choose a quantum state labelled by $(\alpha,\beta)$, and then use these labels to determine the eigenvalues $\lambda=\lambda_f(\alpha,\beta)$ of $Weyl$ as in (\ref{DOIT6}) for Type D spacetimes, or (\ref{OFFTHATFOUR}) for Type I spacetimes.  Next, compute $Weyl$ in the $SO(3,C)$ frame where the Gauss' law constraint is satisfied.  The Gauss' law constraint is a condition on the CDJ matrix

\begin{eqnarray}
\label{GAUSSLAW}
\textbf{w}_e\{(\lambda_f)_{\alpha,\beta}(e^{-\vec{\theta}\cdot{T}})_{fa}(e^{-\vec{\theta}\cdot{T}})_{fe}\}=0,
\end{eqnarray}

\noindent
which reduces to a condition on the three complex $SO(3,C)$ rotation angles $\vec{\theta}\equiv(\theta^1,\theta^2,\theta^3)$.  On solutions to (\ref{GAUSSLAW}) the rotation angles are 
given by $\vec{\theta}=\theta[\vec{\lambda};A]=\theta_{\alpha,\beta}[A]$, which is labelled by the labels $(\alpha,\beta)$ and by the choice of connection $A^a_i$,\footnote{For the purposes of the Gauss' law constraint one may regard the connection $A^a_i$ as a derived quantity from the magnetic field $B^i_a$.  The latter is is turn derived from the vector fields $\textbf{v}_a=B^i_a\partial_i$ tangent to three congruences of integral curves $\vec{\gamma}$ which fill 3-space $\Sigma$.  As shown in papers VI, VII and VIII we regard the integral curves $\vec{\gamma}$ as fundamental, with the magnetic field $B^i_a$ derived upon making a choice of coordinates $x^i$.  In this sense the coordinate-dependent information in the PNDS resides within $\vec{\gamma}$.} and $Weyl$ is given by

\begin{eqnarray}
\label{COMPUTEIT}
\psi_{ae}=(\psi_{\alpha,\beta}[\vec{\gamma}])_{ae}=(e^{\vec{\theta}_{\alpha,\beta}[\vec{\gamma}]\cdot{T}})_{af}(\lambda_f)_{\alpha,\beta}(e^{-\vec{\theta}_{\alpha,\beta}[\vec{\gamma}]\cdot{T}})_{fe}.
\end{eqnarray}

\noindent
One then constructs the Weyl scalars from the elements of (\ref{COMPUTEIT}), of which there should be five

\begin{eqnarray}
\label{COMPUTEIT1}
\Psi_I=(\Psi_{\alpha,\beta}[\vec{\gamma}])_I=T_I^{ae}(\psi_{\alpha,\beta}[\vec{\gamma}])_{ae}.
\end{eqnarray}

\noindent

\newpage

\section{Instanton representation momentum space}

\noindent
We have proven the existence of globally holonomic configuration space variables in the full theory for the instanton representation, which correspond to the existence of quantizable confiruations.  The next thing is to show explicitly which observable characteristics of spacetime correspond to the quantizable degrees of freedom.  The basic momentum space variables are the densitized eigenvalues of the CDJ matrix $\Pi_f=\lambda_f(\hbox{det}A)$, and the undensitized versions $\lambda_f$ directly encode the algebraic classifications of the spacetimes via the invariants $I$ and $J$.  These invariants in turn fix the principal null directions of spacetime.  Therefore in a sense, the quantization of the instanton representation should correspond to a quantization of the principal null directions, and more fundamentally a quantization of the algebraic classification of spacetime.  The degrees of freedom designated for quantization are given by

\begin{eqnarray}
\label{QUUU}
\Psi_{ae}=\bigl(\delta_{ae}\lambda+\alpha(e^2)_{ae}+\beta(e^3_{ae})\bigr)(\hbox{det}A)^{-1},
\end{eqnarray}

\noindent
which we will relate directly to these classifications.  We will treat all cases covering first the nondegenerate spacetimes, where $\Psi_{ae}$ has three independent eigenvectors.  The eigenvalues of such spacetimes must satisfy the Hamiltonian constraint

\begin{eqnarray}
\label{DOIT}
{1 \over \lambda}+{1 \over {\lambda+\alpha}}+{1 \over {\lambda+\beta}}=3\varphi,
\end{eqnarray}

\noindent
irrespective of the degeneracy of the eigenvalues, where $\varphi=-{\Lambda \over 3}(\hbox{det}A)^{-1}$.

\subsection{Type O spacetimes}

We will first consider spacetimes where the CDJ matrix has three linearly independent eigenvectors, with all three eigenvalues equal $\lambda_1=\lambda_2=\lambda_3$.  These are spacetimes of algebraic type O, which include DeSitter spacetime.  The eigenvalues for $Weyl$, the self-dual Weyl curvature $\psi_{ae}$, as well as the Weyl scalars $\Psi_{\alpha}$, are zero for this case

\begin{displaymath}
\psi_{ae}=
\left(\begin{array}{ccc}
0 & 0 & 0\\
0 & 0 & 0\\
0 & 0 & 0\\
\end{array}\right)
;~~\Psi_{\alpha}=
\left(\begin{array}{c}
0\\
0\\
0\\
0\\
0\\
\end{array}\right)
.
\end{displaymath}

\noindent
The CDJ matrix is given by

\begin{displaymath}
\Psi_{ae}=
\left(\begin{array}{ccc}
\lambda & 0 & 0\\
0 & \lambda+\alpha & 0\\
0 & 0 & \lambda+\beta\\
\end{array}\right)
=
\left(\begin{array}{ccc}
\varphi^{-1} & 0 & 0\\
0 & \varphi^{-1} & 0\\
0 & 0 & \varphi^{-1}\\
\end{array}\right)
,
\end{displaymath}

\noindent
which implies that the elements of the deviation from isotropy vanish

\begin{eqnarray}
\label{THENWE}
\alpha=\beta=0.
\end{eqnarray}

\noindent
Since one would like to be able to deduce the properties of the spacetime directly from the state we have, one may alternatively start from the Hamiltonian constraint, which implies that

\begin{eqnarray}
\label{DOIT1}
{3 \over {\lambda_{0,0}}}=3\varphi\longrightarrow\lambda_{0,0}={1 \over \varphi}.
\end{eqnarray}

\noindent
Hence we have that

\begin{eqnarray}
\label{DOIT2}
\lambda_1=\lambda_2=\lambda_3=0;~~I=J=0.
\end{eqnarray}

\subsection{Type D spacetimes}

\noindent
Spacetimes with three linearly independent eigenvectors where there are two independent eigenvalues $\lambda_1=\lambda_2\neq\lambda_3$ for $Weyl$, are of algebraic type D.  The eigenvalues are related by

\begin{displaymath}
\psi_{ae}=
\left(\begin{array}{ccc}
-2\lambda_1 & 0 & 0\\
0 & \lambda_1 & 0\\
0 & 0 & \lambda_1\\
\end{array}\right)
;~~\Psi_{\alpha}=\lambda_1
\left(\begin{array}{c}
0\\
0\\
1\\
0\\
0\\
\end{array}\right)
.
\end{displaymath}

\noindent
For $\alpha=\beta$ the CDJ matrix is given by

\begin{displaymath}
\Psi_{ae}=
\left(\begin{array}{ccc}
\lambda & 0 & 0\\
0 & \lambda+\alpha & 0\\
0 & 0 & \lambda+\beta\\
\end{array}\right)
=
\left(\begin{array}{ccc}
{1 \over {\varphi-2\lambda_1}} & 0 & 0\\
0 & {1 \over {\varphi+\lambda_1}} & 0\\
0 & 0 & {1 \over {\varphi+\lambda_1}}\\
\end{array}\right)
,
\end{displaymath}

\noindent
whence one reads off that

\begin{eqnarray}
\label{OFFTHAT}
\alpha=\beta={1 \over {\varphi+\lambda_1}}-{1 \over {\varphi-2\lambda_1}}.
\end{eqnarray}

\noindent
Since one would like to use the quantum state to make predictions about the spacetime properties directly from the labels of the quantum state, one alternatively may start from $\alpha$ and use the Hamiltonian constraint

\begin{eqnarray}
\label{DOIT3}
{1 \over \lambda}+{2 \over {\lambda+\alpha}}=3\varphi.
\end{eqnarray}

\noindent
This leads to the quadratic equation

\begin{eqnarray}
\label{DOIT4}
3\varphi\lambda^2+3(\alpha\varphi-1)\lambda-\alpha=0
\end{eqnarray}

\noindent
with roots 

\begin{eqnarray}
\label{DOIT5}
\lambda_{\alpha,\alpha}(\varphi)={1 \over {2\varphi}}\Bigl(1-\alpha\varphi\pm\sqrt{(\alpha\varphi)^2-{2 \over 3}(\alpha\varphi)+1}\Bigr).
\end{eqnarray}

\noindent
The roots of $Weyl$ are then given by

\begin{eqnarray}
\label{DOIT6}
\lambda_3=-\varphi+{1 \over {\lambda_{\alpha,\alpha}(\varphi)}};\nonumber\\
\lambda_1=\lambda_2=-\varphi+{1 \over {\alpha+\lambda_{\alpha,\alpha}(\varphi)}},
\end{eqnarray}

\noindent
from which one may directly compute the radiation invariants

\begin{eqnarray}
\label{DOIT7}
I=I(\alpha,\varphi)={1 \over 2}\biggl[\Bigl(-\varphi+{1 \over {\lambda_{\alpha,\alpha}(\varphi)}}\Bigr)^2+2\Bigl(-\varphi+{1 \over {\alpha+\lambda_{\alpha,\alpha}(\varphi)}}\Bigr)^2\biggr];\nonumber\\
J=J(\alpha,\varphi)={1 \over 6}\biggl[\Bigl(-\varphi+{1 \over {\lambda_{\alpha,\alpha}(\varphi)}}\Bigr)^3+2\Bigl(-\varphi+{1 \over {\alpha+\lambda_{\alpha,\alpha}(\varphi)}}\Bigr)^3\biggr].
\end{eqnarray}

\noindent
as a function of the state label $\alpha$.

For $\alpha\neq\beta=0$ the CDJ matrix is given by

\begin{displaymath}
\Psi_{ae}=
\left(\begin{array}{ccc}
\lambda & 0 & 0\\
0 & \lambda & 0\\
0 & 0 & \lambda+\alpha\\
\end{array}\right)
=
\left(\begin{array}{ccc}
{1 \over {\varphi+\lambda_1}} & 0 & 0\\
0 & {1 \over {\varphi+\lambda_1}} & 0\\
0 & 0 & {1 \over {\varphi-2\lambda_1}}\\
\end{array}\right)
,
\end{displaymath}

\noindent
whence one reads off that

\begin{eqnarray}
\label{OFFTHAT1}
\alpha={1 \over {\varphi-2\lambda_1}}-{1 \over {\varphi+\lambda_1}}.
\end{eqnarray}

\noindent
The Hamiltonian constraint in this case is given by

\begin{eqnarray}
\label{DOIT31}
{2 \over \lambda}+{1 \over {\lambda+\alpha}}=3\varphi,
\end{eqnarray}

\noindent
which leads to the quadratic equation

\begin{eqnarray}
\label{DOIT4}
3\varphi\lambda^2+3(\alpha\varphi-1)\lambda-2\alpha=0
\end{eqnarray}

\noindent
with roots 

\begin{eqnarray}
\label{DOIT5}
\lambda_{\alpha,0}(\varphi)={1 \over {2\varphi}}\Bigl(1-\alpha\varphi\pm\sqrt{(\alpha\varphi)^2+{2 \over 3}(\alpha\varphi)+1}\Bigr).
\end{eqnarray}

\noindent
The roots of $Weyl$ are then given by

\begin{eqnarray}
\label{DOIT6}
\lambda_1=\lambda_2=-\varphi+{1 \over {\lambda_{\alpha,0}(\varphi)}};\nonumber\\
\lambda_3=-\varphi+{1 \over {\alpha+\lambda_{\alpha,0}(\varphi)}},
\end{eqnarray}

\noindent
from which one may directly compute the radiation invariants

\begin{eqnarray}
\label{DOIT7}
I=I(\alpha,\varphi)={1 \over 2}\biggl[2\Bigl(-\varphi+{1 \over {\lambda_{\alpha,0}(\varphi)}}\Bigr)^2+\Bigl(-\varphi+{1 \over {\alpha+\lambda_{\alpha,0}(\varphi)}}\Bigr)^2\biggr];\nonumber\\
J=J(\alpha,\varphi)={1 \over 6}\biggl[2\Bigl(-\varphi+{1 \over {\lambda_{\alpha,0}(\varphi)}}\Bigr)^3+\Bigl(-\varphi+{1 \over {\alpha+\lambda_{\alpha,0}(\varphi)}}\Bigr)^3\biggr].
\end{eqnarray}

\noindent
as a function of the state label $\alpha$.

\subsection{Type I spacetimes}

\noindent
Type I spacetimes are algebraically general, and posses three independent eigenvalues $\lambda_1\neq\lambda_2\neq\lambda_3$ for $Weyl$ and three linearly independent eigenvectors.  The eigenvalues are related by

\begin{displaymath}
\psi_{ae}=
\left(\begin{array}{ccc}
-\lambda_1-\lambda_2 & 0 & 0\\
0 & \lambda_1 & 0\\
0 & 0 & \lambda_2\\
\end{array}\right)
;~~\Psi_{\alpha}={{\lambda_1} \over 2}
\left(\begin{array}{c}
1\\
0\\
1\\
0\\
1\\
\end{array}\right)
+{{\lambda_2} \over 2}
\left(\begin{array}{c}
-1\\
0\\
1\\
0\\
-1\\
\end{array}\right)
.
\end{displaymath}

\noindent
The CDJ matrix is given by

\begin{displaymath}
\Psi_{ae}=
\left(\begin{array}{ccc}
\lambda & 0 & 0\\
0 & \lambda+\alpha & 0\\
0 & 0 & \lambda+\beta\\
\end{array}\right)
=
\left(\begin{array}{ccc}
{1 \over {\varphi-\lambda_1-\lambda_2}} & 0 & 0\\
0 & {1 \over {\varphi+\lambda_1}} & 0\\
0 & 0 & {1 \over {\varphi+\lambda_2}}\\
\end{array}\right)
,
\end{displaymath}

\noindent
whence one reads off that

\begin{eqnarray}
\label{OFFTHATONE}
\alpha={1 \over {\varphi+\lambda_1}}-{1 \over {\varphi-\lambda_1-\lambda_2}};~~
\beta={1 \over {\varphi+\lambda_2}}-{1 \over {\varphi-\lambda_1-\lambda_2}}.
\end{eqnarray}

\noindent
We would like to read off the properties of the spacetime directly from the quantum state, for any pair $(\alpha,\beta)$.  Hence the Hamiltonian constraint reads

\begin{eqnarray}
\label{OFFTHATTWO}
{1 \over \lambda}+{1 \over {\lambda+\alpha}}+{1 \over {\lambda+\beta}}=3\varphi.
\end{eqnarray}

\noindent
Equation (\ref{OFFTHATTWO}) leads to the cubic equation

\begin{eqnarray}
\label{OFFTHATTHREE}
\lambda^3+\Bigl(\alpha+\beta-{1 \over {3\varphi}}\Bigr)\lambda^2
+\Bigl(\alpha\beta-{2 \over {3\varphi}}(\alpha+\beta)\Bigr)\lambda-{{\alpha\beta} \over {3\varphi}}=0,
\end{eqnarray}

\noindent
with three roots $\lambda_{\alpha,\beta}$ labelled by $\alpha$ and $\beta$.\footnote{We do not display the explicit solution for the roots here, but they can be found by the method of Cardano.}  Then one solves for the eigenvalues of $Weyl$, given by

\begin{eqnarray}
\label{OFFTHATFOUR}
\lambda_3=-\varphi+{1 \over {\lambda_{\alpha,\beta}(\varphi)}};\nonumber\\
\lambda_1=-\varphi+{1 \over {\alpha+\lambda_{\alpha,\beta}(\varphi)}};\nonumber\\
\lambda_2=-\varphi+{1 \over {\beta+\lambda_{\alpha,\beta}(\varphi)}},
\end{eqnarray}

\noindent
from which one can directly compute the invariants

\begin{eqnarray}
\label{OFFTHATFIVE}
I=I(\alpha,\beta)={1 \over 2}\biggl[\Bigl(-\varphi+{1 \over {\lambda_{\alpha,\beta}(\varphi)}}\Bigr)^2;\nonumber\\
+\Bigl(-\varphi+{1 \over {\alpha+\lambda_{\alpha,\beta}(\varphi)}}\Bigr)^2+\Bigl(-\varphi+{1 \over {\beta+\lambda_{\alpha,\beta}(\varphi)}}\Bigr)^2\biggr];\nonumber\\
J=J(\alpha,\beta)={1 \over 6}\biggl[\Bigl(-\varphi+{1 \over {\lambda_{\alpha,\beta}(\varphi)}}\Bigr)^3\nonumber\\
+\Bigl(-\varphi+{1 \over {\alpha+\lambda_{\alpha,\beta}(\varphi)}}\Bigr)^3+\Bigl(-\varphi+{1 \over {\beta+\lambda_{\alpha,\beta}(\varphi)}}\Bigr)^3\biggr]
\end{eqnarray}

\noindent
as a function of the state labels.

\subsection{Degenerate spacetimes}

\noindent
For the nondegenerate cases it was straightforward to identify the quantizable degrees of freedom because $\Psi_{ae}$ could be taken to be already in diagonal form, known as the intrinsic $SO(3,C)$ frame.  We will see for the degenerate cases that a intrinsic $SO(3,C)$ frame cannot be defined since $\Psi_{ae}$ is not diagonlizable.  This is due to the fact that the number of linearly independent eigenvectors is less than the rank of the 
matrix.\footnote{One could determine the diagonalizable subspace of $\Psi_{ae}$, and attampt to perform a quantization restricted to this subspace.  We relegate such treatments for future study.}  We will now consider each case in turn.\par
\indent
Type N spacetimes have one independent eigenvalue $\lambda_1=\lambda_2=0$ for $Weyl$, and two linearly independent eigenvectors.  For this case $Weyl$ is given by

\begin{displaymath}
\psi_{ae}=k
\left(\begin{array}{ccc}
0 & 0 & 0\\
0 & 1 & i\\
0 & i & -1\\
\end{array}\right)
;~~\Psi_{\alpha}=
\left(\begin{array}{c}
2\\
0\\
0\\
0\\
0\\
\end{array}\right)
\end{displaymath}

\noindent
for some numerical constant $k$.  The CDJ matrix is given by

\begin{displaymath}
\Psi_{ae}=
\left(\begin{array}{ccc}
{1 \over \varphi} & 0 & 0\\
0 & {1 \over \varphi}-{k \over {\varphi^2}} & -i{k \over {\varphi^2}}\\
0 & -i{k \over {\varphi^2}} & {1 \over \varphi}+{k \over {\varphi^2}}\\
\end{array}\right)
.
\end{displaymath}

\noindent
The attempt to diagonalize this leads to the matrix

\begin{displaymath}
\Psi_{ae}=
\left(\begin{array}{ccc}
\lambda & 0 & 0\\
0 & \lambda+\alpha & 0\\
0 & 0 & \lambda+\beta\\
\end{array}\right)
=
\left(\begin{array}{ccc}
{1 \over \varphi} & 0 & 0\\
0 & {1 \over \varphi}-{k \over {\varphi^2}}e^{-2i\phi} & 0\\
0 & 0 & {1 \over \varphi}+{k \over {\varphi^2}}e^{-2i\phi}\\
\end{array}\right)
;~~\hbox{tan}{2\phi}=-i.
\end{displaymath}

\noindent
which is ill-defined.  One may nevertheless attempt to read off $\alpha$ and $\beta$, given by

\begin{eqnarray}
\label{ILLDEF}
\alpha=-{k \over {\varphi^2}}e^{-2i\phi};~~\beta={k \over {\varphi^2}}e^{-2i\phi}
\end{eqnarray}

\noindent
which are also ill-defined.\par 
\indent
Type III spacetimes have one independent eigenvalue for $Weyl$, and one linearly independent eigenvector.  The eigenvalue is zero for this case

\begin{displaymath}
\psi_{ae}=k
\left(\begin{array}{ccc}
0 & 1 & 0\\
1 & 0 & i\\
0 & i & 0\\
\end{array}\right)
;~~\Psi_{\alpha}=k
\left(\begin{array}{c}
1\\
-{i \over 2}\\
0\\
-{i \over 2}\\
-1\\
\end{array}\right)
.
\end{displaymath}

\noindent
for some numerical constant $k$.  The CDJ matrix is given by

\begin{displaymath}
\Psi_{ae}=
\left(\begin{array}{ccc}
{1 \over \varphi}+{{k^2} \over {\varphi^3}} & -{k \over {\varphi^2}} & i{{k^2} \over {\varphi^3}}\\
-{k \over {\varphi^2}} & {1 \over \varphi} & i{k \over {\varphi^2}}\\
i{{k^2} \over {\varphi^3}} & i{k \over {\varphi^2}} & {1 \over \varphi}-{{k^2} \over {\varphi^3}}\\
\end{array}\right)
.
\end{displaymath}

\noindent
Type N spacetimes have two independent eigenvalues for $Weyl$, and two linearly independent eigenvectors.  The eigenvalue is zero for this case

\begin{displaymath}
\psi_{ae}=k
\left(\begin{array}{ccc}
0 & 0 & 0\\
0 & 1 & i\\
0 & i & -1\\
\end{array}\right)
+\lambda_2
\left(\begin{array}{ccc}
-2 & 0 & 0\\
0 & 1 & 0\\
0 & 0 & 1\\
\end{array}\right)
;~~\Psi_{\alpha}=
\left(\begin{array}{c}
2\\
0\\
0\\
0\\
0\\
\end{array}\right)
+\lambda_2
\left(\begin{array}{c}
0\\
0\\
1\\
0\\
0\\
\end{array}\right)
.
\end{displaymath}

\noindent
for some numerical constant $k$.  The CDJ matrix is given by

\begin{displaymath}
\Psi_{ae}=(\varphi+\lambda_2)^{-2}
\left(\begin{array}{ccc}
{{(\varphi+\lambda_2)^2} \over {\varphi-2\lambda_2}} & 0 & 0\\
0 & \varphi+\lambda_2+k & -ik\\
0 & ik & \varphi+\lambda_2-k\\
\end{array}\right)
.
\end{displaymath}

\noindent
Attempts to diagonalize this yield

\begin{displaymath}
\Psi_{ae}=
\left(\begin{array}{ccc}
\lambda & 0 & 0\\
0 & \lambda+\alpha & 0\\
0 & 0 & \lambda+\beta\\
\end{array}\right)
\end{displaymath}

\begin{displaymath}
=
(\varphi+\lambda_2)^{-2}
\left(\begin{array}{ccc}
{{(\varphi+\lambda_2)^2} \over {\varphi-2\lambda_2}} & 0 & 0\\
0 & \varphi+\lambda_2+ke^{-2i\theta} & 0\\
0 & 0 & \varphi+\lambda_2-ke^{-2i\theta}\\
\end{array}\right)
;~~\hbox{tan}(2\theta)=-i.
\end{displaymath}

\noindent
From this one may read off $\alpha$ and $\beta$, given by

\begin{eqnarray}
\label{ILLDEFINE}
\alpha=(\varphi+\lambda_2)^{-2}\Bigl(\varphi+\lambda_2+ke^{-2i\theta}-{{(\varphi+\lambda_2)^2} \over {\varphi-2\lambda_2}}\Bigr);\nonumber\\
\beta=(\varphi+\lambda_2)^{-2}\Bigl(\varphi+\lambda_2-ke^{-2i\theta}-{{(\varphi+\lambda_2)^2} \over {\varphi-2\lambda_2}}\Bigr).
\end{eqnarray}

\noindent
Equation (\ref{ILLDEFINE}) are ill-defined, as with the rest of the degenerate cases.

\newpage

\newpage

\section{Conclusion}

The main result of this paper is as follows.  We have shown that there exists a natural canonical structure associated with the Petrov classification of spacetime for spacetimes of Petrov type $I$, $D$ and $O$.  This canonical structure corresponds to the kinematic phase space of the instanton representation of Plebanski gravity, the phase space at the level of implementation of the kinematic constraints.  The Hamiltonian constraint at this level fixes the algebraic classification of the spacetime through the eigenvalues of the CDJ matrix $\Psi_{ae}$, and enables one to quantize the theory in an intrinsic $SO(3,C)$ frame.  The aformentioned canonical structure provides globally holonomic coordinates on the instanton representation configuration space, which requires the use of a denstized version of the CDJ matrix as the basic momentum space variable.  Once the elements of $\Psi_{ae}$ have been densitized by the Chern--Simons Lagrangian, then the implementation of the initial value constraints yields a kinematic phase space space where this canonical structure is realized.  In essense, the instanton representation admits a canonical structure which allows for the possibility to quantize gravity and obtain a classical limit in terms of the Petrov classification of spacetime, which is in principle directly measurable in this limit.  The next step is the construction of a Hilbert space for such a quantum theory, which is the topic of Papers XV and XVIII.

\newpage

\section{Appendix A. Hamiltonian constraint in polynomial form}

\noindent
While the Hamiltonian constraint is nonpolynomial in the instanton representation, we will see that it is convenient to extract the polynomial part when quantizing the theory.  The smeared form of the Hamiltonian consraint at the classical level can be written

\begin{eqnarray}
\label{FINAL18}
H[N]=\int_{\Sigma}d^3xN(\hbox{det}B)^{1/2}\sqrt{\hbox{det}\Psi}\Bigl(\Lambda+\hbox{tr}\Psi^{-1}\Bigr)=\int_{\Sigma}d^3xN\sqrt{{{\hbox{det}B} \over {\hbox{det}\Psi}}}\Bigl({1 \over 2}Var\Psi+\Lambda\hbox{det}\Psi\Bigr).
\end{eqnarray}

\noindent
where we have used the characteristic equation

\begin{eqnarray}
\label{CHARACTERISTICEQ}
\hbox{tr}M^{-1}={{Var{M}} \over {2\hbox{det}M}}
\end{eqnarray}

\noindent
for a three by three matrix $M$, where $Var{M}=(\hbox{tr}M)^2-\hbox{tr}M^2$.\par
\indent
Since we are restricting to nondegenerate configurations $\hbox{det}B\neq{0}$ and $\hbox{det}\Psi\neq{0}$, we can then focus on the part

\begin{eqnarray}
\label{FINAL18}
{1 \over 2}Var\Psi+\Lambda\hbox{det}\Psi,
\end{eqnarray}

\noindent
where $Var\Psi=(\hbox{tr}\Psi)^2-\hbox{tr}\Psi^2$.\par
\indent
The CDJ matrix $\Psi_{ae}$ can be parametrized by its symmetric and antisymmetric parts $\lambda_{ae}$ and $a_{ae}$ such that

\begin{eqnarray}
\label{FINAL19}
\Psi_{ae}=\lambda_{ae}+a_{ae}=\lambda_{ae}+\epsilon_{aed}\psi^d
\end{eqnarray}

\noindent
for some arbitrary $SU(2)_{-}$ valued 3-vector $\psi^d$.  The ingredients of the Hamiltonian constraint are then given by

\begin{eqnarray}
\label{FINAL20}
\hbox{det}\Psi={1 \over 6}\epsilon_{abc}\epsilon_{efg}(\lambda_{ae}+\epsilon_{aed_1}\psi^{d_1})(\lambda_{bf}+\epsilon_{bfd_2}\psi^{d_2})(\lambda_{cg}+\epsilon_{cgd_3}\psi^{d_3})\nonumber\\
=\hbox{det}(\lambda_{ae})+\hbox{det}(\epsilon_{aed}\lambda^d)
+{1 \over 2}\epsilon_{abc}\epsilon_{efg}\Bigl(\epsilon_{cgd}\lambda_{ae}\lambda_{bf}\psi^d+\epsilon_{bfd}\epsilon_{cgd^{\prime}}\lambda_{ae}\psi^d\psi^{d^{\prime}}\Bigr).
\end{eqnarray}

\noindent
Using the fact that the determinant of an antisymmetric matrix of odd rank vanishes, and the annihilation of symmetric on antisymmetric quantities, we end up with

\begin{eqnarray}
\label{FINAL21}
\hbox{det}\Psi=\hbox{det}\lambda+{1 \over 2}(\epsilon_{abc}\epsilon_{fbd})(\epsilon_{efg}\epsilon_{ed^{\prime}g})\psi^d\psi^{d^{\prime}}\lambda_{ae}\nonumber\\
=\hbox{det}\lambda+{1 \over 2}\lambda_{ae}(\delta_{af}\delta_{cd}-\delta_{ad}\delta_{cf})(\delta_{ec}\delta_{fd^{\prime}}-\delta_{ed^{\prime}}\delta_{fc})\psi^d\psi^{d^{\prime}}\nonumber\\
=\hbox{det}\lambda+\lambda_{fg}\psi^f\psi^g
\end{eqnarray}

\noindent
where we have made use of epsilon symbol identities.  Likewise, one may compute the variance

\begin{eqnarray}
\label{FINAL22}
Var\Psi=(\hbox{tr}\lambda)^2-(\lambda_{ae}+\epsilon_{aed}\psi^d)(\lambda_{ea}-\epsilon_{aed^{\prime}}\psi^{d^{\prime}})=Var\lambda-2\delta_{fg}\psi_f\psi_g.
\end{eqnarray}

\noindent
The Hamiltonian constraint (\ref{FINAL18}) then is given by

\begin{eqnarray}
\label{FINAL23}
H=Var\lambda+\Lambda\hbox{det}\lambda+(\lambda_{fg}-2\delta_{fg})\psi^f\psi^g.
\end{eqnarray}

\noindent
The symmetric part of the $CDJ$ matrix $\Psi_{ae}$ can be re-written as a complex orthogonal $(SO(3,C)$ transformation parametrized by three complex angles $O=e^{\vec{\theta}\cdot{T}}$, 
where $\vec{\theta}\equiv(\theta^1,\theta^2,\theta^3)$ and $T\equiv(T_1,T_2,T_3)$ are the generators of the $so(3,c)$ algebra in the adjoint representation.\footnote{This has the interpretation of a Lorentz transformation into a new frame, where $Im[\vec{\theta}]$ and $Re[\vec{\theta}]$ represent rotations and boosts respectively.}  This is given by

\begin{eqnarray}
\label{FINAL24}
\lambda_{ae}=(e^{\theta\cdot{T}})_{af}\lambda_f(e^{\theta\cdot{T}})^T_{fe}.
\end{eqnarray}

\noindent
where $\lambda_f\equiv(\lambda_1,\lambda_2,\lambda_2)$ are the eigenvalues of the symmetric part.  Since the first two terms of the Hamiltonian 
constraint (\ref{FINAL23}) depend upon the invariants of $\lambda_{ae}$, then the matrix $O_{ae}$ cancels out and we obtain

\begin{eqnarray}
\label{FINAL25}
H=2(\lambda_1\lambda_2+\lambda_2\lambda_3+\lambda_3\lambda_1)+\Lambda\lambda_1\lambda_2\lambda_3+(\lambda_f-2){\psi^{\prime}}^f{\psi^{\prime}}^f,
\end{eqnarray}

\noindent
where $\psi^{\prime}_f=(e^{\theta\cdot{T}})_{fd}\psi^d$ is the Lorentz transformation of $\psi^d$ into a new Lorentz frame.\par
\indent
We would rather like to interpret the Hamiltonian constraint as being independent of the Lorentz frame, and consider the angles $\vec{\theta}$ as not being independent physical degrees of freedom.  The most direct way to do this is to require that $\psi^d=0$, which imples that the antisymmetric part of the $CDJ$ matrix $\Psi_{[ae]}$ vanish.

\end{document}